\documentclass[PAPER, atlasdraft=false, cernpreprint=true, USenglish, orcidlogo, texmf, orcidlogo]{atlasdoc}
 
\usepackage{atlaspackage}
\usepackage{atlasbiblatex}
\usepackage{multirow}
\usepackage{atlasphysics}

\graphicspath{{logos/}{figures/}}
 
\usepackage{ANA-STDM-2020-01-PAPER-defs}

\AtlasTitle{Studies of the energy dependence of diboson polarization fractions and the Radiation Amplitude Zero effect in $WZ$ production with the ATLAS detector}

\AtlasAbstract{
This Letter presents the first study of the energy-dependence of diboson polarization fractions in $WZ \rightarrow \ell\nu \ell'\ell'~(\ell, \ell'=e, \mu)$ production.
The data set used corresponds to an integrated luminosity of 140 fb$^{-1}$ of proton-proton collisions at a center-of-mass energy of 13 TeV recorded by the ATLAS detector.
Two fiducial regions with an enhanced presence of events featuring two longitudinally-polarized bosons are defined.
A non-zero fraction of events with two longitudinally-polarized bosons is measured with an observed significance of 5.2 standard deviations in the region with $100<p_T^Z\leq200$ GeV and 1.6 standard deviations in the region with $p_T^Z>200$ GeV, where $p_T^Z$ is the transverse momentum of the $Z$ boson.
This Letter also reports the first study of the Radiation Amplitude Zero effect. Events with two transversely-polarized bosons are analyzed for the $\Delta Y(\ell_W Z)$ and $\Delta Y(WZ)$ distributions defined respectively as the rapidity difference between the lepton from the $W$ boson decay and the $Z$ boson and the rapidity difference between the $W$ boson and the $Z$ boson. Significant suppression of events near zero is observed in both distributions. Unfolded $\Delta Y(\ell_W Z)$ and $\Delta Y(WZ)$ distributions are also measured and compared to theoretical predictions.
 
}

\AtlasRefCode{STDM-2020-01}
 
\PreprintIdNumber{CERN-EP-2024-045}

\AtlasJournal{\PRL}
\AtlasJournalRef{\PRL 133 (2024) 101802}
\AtlasDOI{10.1103/PhysRevLett.133.101802}

\hypersetup{pdftitle={ATLAS document},pdfauthor={The ATLAS Collaboration}}
 
\begin{document}

\maketitle
In the Standard Model (SM) of particle physics, the longitudinal polarization components of the $W$ and $Z$ bosons are generated by the Goldstone bosons from electroweak symmetry breaking via the Higgs mechanism. Physics beyond the SM could cause different effects on the polarization of $W$ (or $Z$) bosons in diboson processes~\cite{Franceschini:2017xkh, Liu:2019vid}.
A sensitive test of the mechanism of electroweak symmetry breaking and gauge symmetry can thus be obtained by studying different polarization states in diboson processes.
This Letter presents the first study of the energy dependence of $WZ$ diboson polarization fractions in two fiducial regions with an enhanced presence of events featuring two longitudinally-polarized bosons. In addition, it reports the first study of the Radiation Amplitude Zero (RAZ) effect in $WZ$ production with two transversely-polarized bosons~\cite{Frixione:1992pj, Baur:1994ia}.
The $WZ$ candidates are reconstructed using leptonic decay modes of the gauge bosons into electrons and muons, $WZ \rightarrow \ell\nu \ell'\ell'$ ($\ell, \ell'=e, \mu$). The data set used corresponds to an integrated luminosity of 140 fb$^{-1}$ of proton-proton collisions at a center-of-mass energy of 13 TeV recorded by the ATLAS detector from 2015 to 2018.
 
At leading-order (LO) in quantum chromodynamics (QCD), $WZ$ production occurs through quark-antiquark interactions in the $s$-, $t$-, and $u$-channels. The dominant helicity amplitude with two transversely-polarized bosons is exactly zero when the scattering angle of the $W$ boson in the $WZ$ rest frame with respect to the incoming antiquark direction approaches $90^\circ$~\cite{Frixione:1992pj, Baur:1994ia}.
This is a direct consequence of the gauge structure in the SM. This RAZ effect leads to a dip around 0 in the $\Delta Y(WZ)$ and $\Delta Y(\ell_WZ)$ distributions, with $\Delta Y(WZ)$ defined as the rapidity difference between the $W$ and $Z$ bosons, and $\Delta Y(\ell_WZ)$ defined as the rapidity difference between the lepton from the $W$ decay and the $Z$ boson.
The RAZ effect has been observed for $W\gamma$~\cite{D0:2008abl, CMS-EWK-11-009, CMS-SMP-20-005} for which it is found that the sensitivity for $W\gamma$ resonances is enhanced in this radiation valley~\cite{Capdevilla:2019zbx}. However, the RAZ effect has not yet been observed for $WZ$ due to the $W$ boson polarizations in $WZ$ production~\cite{Bilchak:1984gv}.
In addition, the next-to-leading order (NLO) QCD corrections dilute the RAZ effect and make it hard to observe experimentally~\cite{Baur:1994aj, Frye:2015rba}.
To reduce jet activity and to increase the significance of the dips, a selection criterion on the transverse momentum of the $WZ$ system (\wzpT) is applied.
 
The diboson polarization fractions $f_{00}$, $f_{TT}$, $f_{0T}$ and $f_{T0}$ as defined in Ref.~\cite{STDM-2022-01} are interpreted as probabilities of correlated polarization states of the $W$ and $Z$ bosons. Here, $00$ ($TT$) indicates that both bosons are longitudinally (transversely) polarized, and $0T$ ($T0$) indicates that the $W$ ($Z$) boson is longitudinally polarized and the $Z$ ($W$) boson is transversely polarized.
The ATLAS Collaboration has measured both single and diboson polarization fractions using inclusive $WZ$ events~\cite{STDM-2022-01}, which are dominated by $TT$ events with low momentum $W$ and $Z$ bosons~\cite{Franceschini:2017xkh, Liu:2019vid,Willenbrock:1987xz}.
This analysis focuses on $WZ$ events with $Z$ bosons required to have high transverse momenta (\zpT).
The combination of high $p_T^Z$ and low $p_T^{WZ}$ significantly reduces the $TT$ contribution and increases $f_{00}$.
As a result, $f_{00}$ increases from $5-7\%$ in the inclusive region to $20-30\%$ in the region with high $p_T^Z$ and low $p_T^{WZ}$~\cite{Huang:2020iya}.
 
The ATLAS detector~\cite{PERF-2007-01} is a multi-purpose particle physics detector with cylindrical geometry\footnote{The ATLAS experiment uses a right-handed coordinate system with its origin at the nominal interaction point (IP) in the center of the detector. The $x$-axis points from the IP to the center of the LHC ring, the $y$-axis points upward, and the $z$-axis is along one of the proton beam directions. Cylindrical coordinates $(r, \phi)$ are used in the transverse plane, $\phi$ being the azimuthal angle around the beam pipe. The pseudorapidity is defined in terms of the polar angle $\theta$ as $\eta = -\ln \tan(\theta/2)$. Transverse momentum ($p_T$) is defined relative to the beam axis and is calculated as $p_T = p \sin \theta$ where $p$ is the momentum. The distance $\Delta R$ is defined as $\Delta R=\sqrt{(\Delta \eta)^2+(\Delta \phi)^2}$.}. It consists of an inner tracking detector (ID) surrounded by a superconducting solenoid providing a 2 T axial magnetic field, sampling electromagnetic (ECAL) and hadronic (HCAL) calorimeters using liquid argon as active material and lead, copper or tungsten as absorber material, and a muon spectrometer (MS) based on three air-core toroidal superconducting magnets. The ATLAS trigger system consists of a hardware-based level-1 trigger followed by a software-based high-level trigger~\cite{TRIG-2016-01}. Events used for this analysis are selected with single-lepton ($e$ or $\mu$) triggers~\cite{TRIG-2016-01, TRIG-2018-05, ATL-DAQ-PUB-2019-001}. An extensive software suite~\cite{ATL-SOFT-PUB-2021-001} is used in the reconstruction and analysis of real and simulated data, in detector operations, and in the trigger and data acquisition systems of the experiment.
 
Electron candidates are reconstructed from energy clusters in the ECAL matched to ID tracks. Electrons are identified using a likelihood function based on shower shape variables in the ECAL, track variables, and the quality of the track-cluster matching. Electrons are required to satisfy a ``medium'' likelihood requirement~\cite{EGAM-2018-01} and have $p_T>15$ GeV and $|\eta|<2.47$ excluding the crack region $1.37<|\eta|<1.52$. Muons are reconstructed by matching the tracks in the MS with tracks in the ID. Muons are required to pass a ``medium'' identification selection~\cite{MUON-2018-03} and have $p_T > 15$ GeV and $|\eta|<2.7$. Electrons and muons must be compatible with the hypothesis that they originate from the primary vertex~\cite{EGAM-2018-01, MUON-2018-03}. They are also required to be isolated from other particles using both calorimeter-cluster and ID-track information~\cite{HIGG-2017-02}.
 
Jets are reconstructed with the anti-$k_t$ algorithm with a radius parameter of 0.4, using a particle-flow procedure~\cite{PERF-2015-09} with clusters of energy deposited in the calorimeter and tracks reconstructed in the ID as inputs. Jets are required to have $p_T > 30$ GeV and $|\eta| < 4.5$. To suppress pile-up jets, jets with $p_T<60$ GeV and $|\eta| < 2.4$ have to pass a requirement on the jet vertex tagger~\cite{PERF-2014-03}.
The missing transverse momentum vector with magnitude denoted by \met, is calculated as the negative vector sum of the transverse momentum of all identified ``hard''
physics objects (electrons, muons, jets), with a contribution from an additional soft term based on ID tracks matched to the primary vertex but not assigned to any of the ``hard'' objects~\cite{PERF-2016-07}. Ambiguities in the identity of reconstructed leptons, jets, and photons are resolved with an overlap-removal procedure described in Ref.~\cite{STDM-2017-22}.
 
Candidate $WZ$ events are selected using triggers that require at least one electron or muon. To ensure the trigger efficiency is well determined, at least one of the candidate leptons in the event must be trigger-matched and have $p_T>25$ GeV for data taken in 2015 or $p_T>27$ GeV for data taken from 2016 to 2018.
Events are required to contain exactly three lepton candidates satisfying the selection criteria described above. To reduce contributions from the $ZZ^{(*)} \rightarrow \ell\ell\ell'\ell'$ process, events with a fourth lepton candidate satisfying looser selection criteria are rejected. For this looser selection, the lepton $p_T$ requirement is reduced to $p_T > 5$ GeV. Electrons are allowed to be reconstructed in the region $1.37 < |\eta| < 1.52$ and ``loose'' identification requirements~\cite{EGAM-2018-01, MUON-2018-03} are used for both the electrons and muons. A less stringent requirement is also applied for electron isolation~\cite{STDM-2022-01}. Candidate events are required to have at least one pair of leptons with the same flavor and opposite charge, with an invariant mass that is consistent with the $Z$ boson pole mass (91.188 GeV \cite{Zyla:2020zbs}) to within 10 GeV. This pair is considered to be the $Z$ boson candidate. If more than one pair can be formed, the pair whose invariant mass is closest to the $Z$ boson pole mass is taken as the $Z$ boson candidate. The remaining third lepton is assigned to the $W$ boson decay. To suppress backgrounds with non-prompt leptons from hadron (including $b$-flavored and $c$-flavored hadrons) decays and jets misidentified as leptons, the $W$ lepton is required to pass tighter isolation requirements and have $p_T>20$ GeV. The transverse mass of the $W$ candidate, defined as $m^{W}_{\textrm T}=\sqrt{2 p_T^\ell \met[1 - \cos \Delta \phi]}$, is required to be greater than 30 GeV, where $\Delta \phi$ is the angle between the third lepton and the \met in the transverse plane, and $p_T^\ell$  the transverse momentum of the third lepton. These selection criteria are used to select inclusive $WZ$ events. The overall acceptance times efficiency after these selections ranges from 40-60\% depending on the channel.

The $W$-mass constraint method is used to calculate the longitudinal component of the neutrino momentum ($p_z(\nu)$). The $\met$ of the event is assumed to come from the neutrino, and $p_z(\nu)$ is estimated by constraining the invariant mass of the third lepton and the neutrino to be the pole mass of the $W$ boson (80.385 GeV \cite{Zyla:2020zbs}). A quadratic equation leads to two solutions. If they are real, the one with the smaller magnitude of $|p_z(\nu)|$ is chosen, otherwise, the real part is chosen~\cite{STDM-2018-03}. The effect of these choices is consistent between data and MC and does not bias our measurements.

For RAZ studies, an additional criterion of $p_T^{WZ}<$ 20, 40, or 70 GeV is applied to define three regions with reduced jet activity in each event. Diboson polarization fractions are measured in two regions enhanced in events with $00$ polarization ($00$-enhanced), defined with two additional criteria applied on the inclusive region: $p_T^{WZ}<70$ GeV, and either $100<p_T^Z\leq200$ GeV or $p_T^Z>200$ GeV.
 
Monte Carlo (MC) simulated samples are used to model the signal $WZ$ process with different polarization states, as well as other physics processes with at least three prompt leptons. Simulated events are processed through the full ATLAS detector simulation based on \textsc{geant4}~\cite{Agostinelli:2002hh, SOFT-2010-01}, and reconstructed with the same algorithms as those used for the data. To simulate pileup effects, additional $pp$ interactions are added to the MC samples in proportion to the mean interactions per bunch crossing occurring in the various data periods.
 
For inclusive $WZ$ production, NLO QCD~\cite{Ohnemus:1991gb, Frixione:1992pj} and NLO electroweak corrections~\cite{Bierweiler:2013dja, Baglio:2013toa, Biedermann:2017oae} have been calculated. Next-to-next-to-leading-order (NNLO) QCD corrections are also known in the on-shell case, including off-shell effects~\cite{Grazzini:2016swo, Grazzini:2017ckn} and combined with electroweak corrections~\cite{Grazzini:2019jkl}. Diboson polarizations in $WZ$ production have been studied at NLO QCD~\cite{Denner:2020eck, Pelliccioli:2023zpd} and NLO QCD+electroweak accuracy~\cite{Baglio:2018rcu, ninh2023enhancing}. However, the automated MC simulation with two polarized vector bosons and full spin correlations is only available at LO in the narrow-width approximation.
In this analysis, polarized $WZ$ events are generated at LO using \textsc{Madgraph\_amc@nlo} 2.7.3~\cite{BuarqueFranzosi:2019boy}. The boson polarizations are defined in the $WZ$ center-of-mass frame. To account for the real part of NLO QCD corrections, events are simulated with no jets and with one jet at LO, and the two samples are merged with \textsc{pythia8}~\cite{Sjostrand:2007gs} using the CKKW-L scheme \cite{Lonnblad:2001iq}. Jets are defined to have $p_T>25$ GeV and $|\eta|<2.5$ at the parton level. In addition, \textsc{pythia8} is used for the simulation of parton showering, hadronization and the underlying event. The NNPDF3.0nlo PDF set~\cite{Ball:2012cx} is used for the parton process generation. Separate MC samples are generated corresponding to the four states $00$, $0T$, $T0$ and $TT$, respectively. For each polarization state, different samples are generated for $p_T^Z>150$ GeV and $p_T^Z \le 150$ GeV in order to increase statistics for events with high $p_T^Z$.
Separate samples are also generated for events with at least one boson decays to $\tau$ lepton and the $\tau$ lepton decays leptonically to an electron or a muon. An inclusive MC sample is created by adding the four polarized samples. These MC sets are referred to as {\textsc{Madgraph}} 0,1j@LO and the inclusive cross section is scaled to the NLO QCD prediction~\cite{Dixon:1999di, Campbell:1999ah, Campbell:2011bn}.
Scale factors are also derived to reweight simulated {\textsc{Madgraph}} 0,1j@LO events to agree with data for three jet multiplicity bins (0 jets, 1 jet, and $\ge 2$ jets) in the inclusive region to account for missing higher-order QCD effects.
 
To cross check the modeling of polarized $WZ$ events, an inclusive $WZ \rightarrow \ell\nu\ell'\ell'$ sample is generated at NLO in QCD with
\textsc{sherpa} 2.2.2~\cite{Sherpa:2019gpd}.
\textsc{pythia8} is used for the modeling of parton shower, hadronization, and underlying event. The inclusive {\textsc{Madgraph}} 0,1j@LO sample is compared to this \textsc{sherpa} sample.
In addition, the {\textsc{Madgraph}} 0,1j@LO sample is also compared to the NLO QCD and NLO QCD+electroweak calculations~\cite{Pelliccioli:2023zpd,ninh2023enhancing, Baglio:2018rcu} for each polarization state.
In general good agreement is observed for polarization fractions and kinematic distributions in both inclusive and $00$-enhanced signal regions.
 
Due to the requirement of three isolated charged leptons in the final state, only about 10\% of events come from background processes.
To estimate background processes with three prompt leptons in the final state, $q\bar{q}$-initiated and $gg$-initiated $ZZ^{(*)}$ processes are simulated using \textsc{sherpa} 2.2.2. It provides a matrix element calculation accurate to NLO in QCD for 0- and 1-jet final states, and LO accuracy for 2- and 3-jet final states. Both $t\bar{t}Z$ and $t\bar{t}W$ processes are generated at NLO in QCD using the \textsc{Madgraph\_amc@NLO} 2.6.5 generator and interfaced with \textsc{pythia~8} for the modelling of the parton shower. Triboson ($VVV$ with $V=W, Z$) events are simulated using the \textsc{sherpa} generator at NLO in QCD with 0 partons and at LO accuracy with 1 and 2 additional partons. These background samples are normalized to cross sections with higher-order corrections applied~\cite{Grazzini:2015hta, Binoth:2008kt, Frederix:2021agh, Denner:2023eti}.
Backgrounds with at least one misidentified lepton (labeled as ``non-prompt background'') are evaluated using a data-driven technique described in Refs.~\cite{SUSY-2013-09, STDM-2022-01}.
 
Instrumental systematic uncertainties are related to the lepton trigger, reconstruction and identification efficiencies~\cite{MUON-2018-03, PERF-2017-01}, lepton isolation criteria~\cite{HIGG-2017-02}, lepton energy (momentum) scale and resolution~\cite{MUON-2018-03, PERF-2017-03}, jet energy scale and resolution~\cite{JETM-2018-05}, jet vertex tagging~\cite{ATLAS-CONF-2014-018, PERF-2016-06}, $b$-jet identification~\cite{FTAG-2018-01}, modeling of \met~\cite{PERF-2016-07} and pile-up, and integrated luminosity~\cite{ATLAS-CONF-2019-021, Avoni:2018iuv}.
 
Theoretical uncertainties associated with the signal and other background processes are evaluated using simulation.
Shape and acceptance uncertainties on the $WZ$ process due to renormalization and factorization scales~\cite{ATL-PHYS-PUB-2016-002}, parton distribution functions~\cite{Butterworth:2015oua},
and parton shower, are also considered.
The normalization uncertainties on the processes included in the ``Prompt'' background category are between $10$--$20$\%~\cite{TOPQ-2018-01, STDM-2017-06, TOPQ-2016-11}. Systematic uncertainties due to the potential mismodeling of NLO QCD and electroweak corrections are estimated by applying reweighting corrections estimated from Ref. \cite{ninh2023enhancing} as a function of different sets of variables: the cosine of the scattering angle of the $W$ boson in the $WZ$ rest frame with respect to the $z$-axis ($\cos \theta_V$) and $\Delta Y(\ell_W Z)$ for NLO QCD corrections, and average $p_T$ of the three charged leptons and $\Delta Y(\ell_W Z)$ for NLO electroweak corrections.
Additionally, uncertainties from the implementation of the NLO electroweak corrections are estimated by taking the difference between the additive and the multiplicative prescription as described in \cite{Grazzini:2019jkl}.
The interference effects among different polarization states are found to be negligible.
The methods used are similar to the ones described in Ref.~\cite{STDM-2021-05}.

\begin{figure}[htbp]
\centering
\subfloat[ ]{\includegraphics[width=0.45\textwidth]{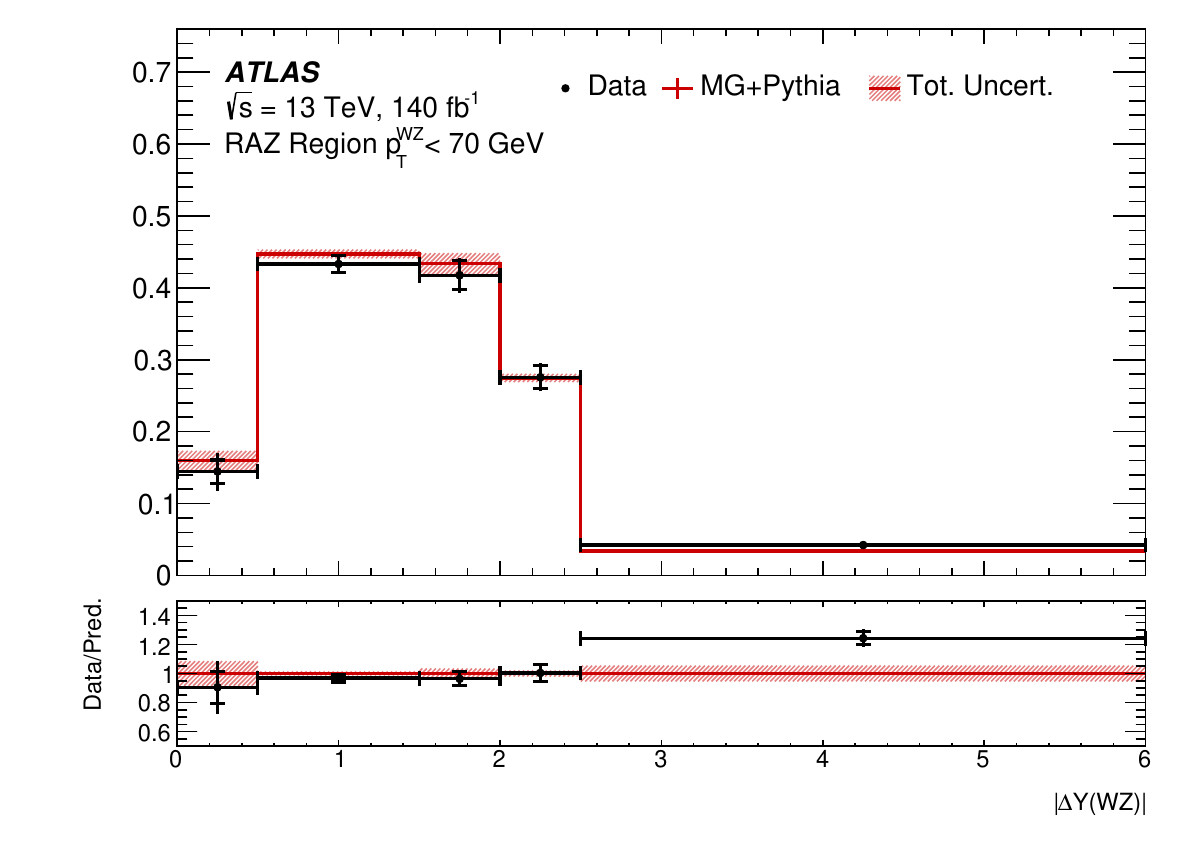}\label{fig:unf}}
\subfloat[ ]{\includegraphics[width=0.45\textwidth]{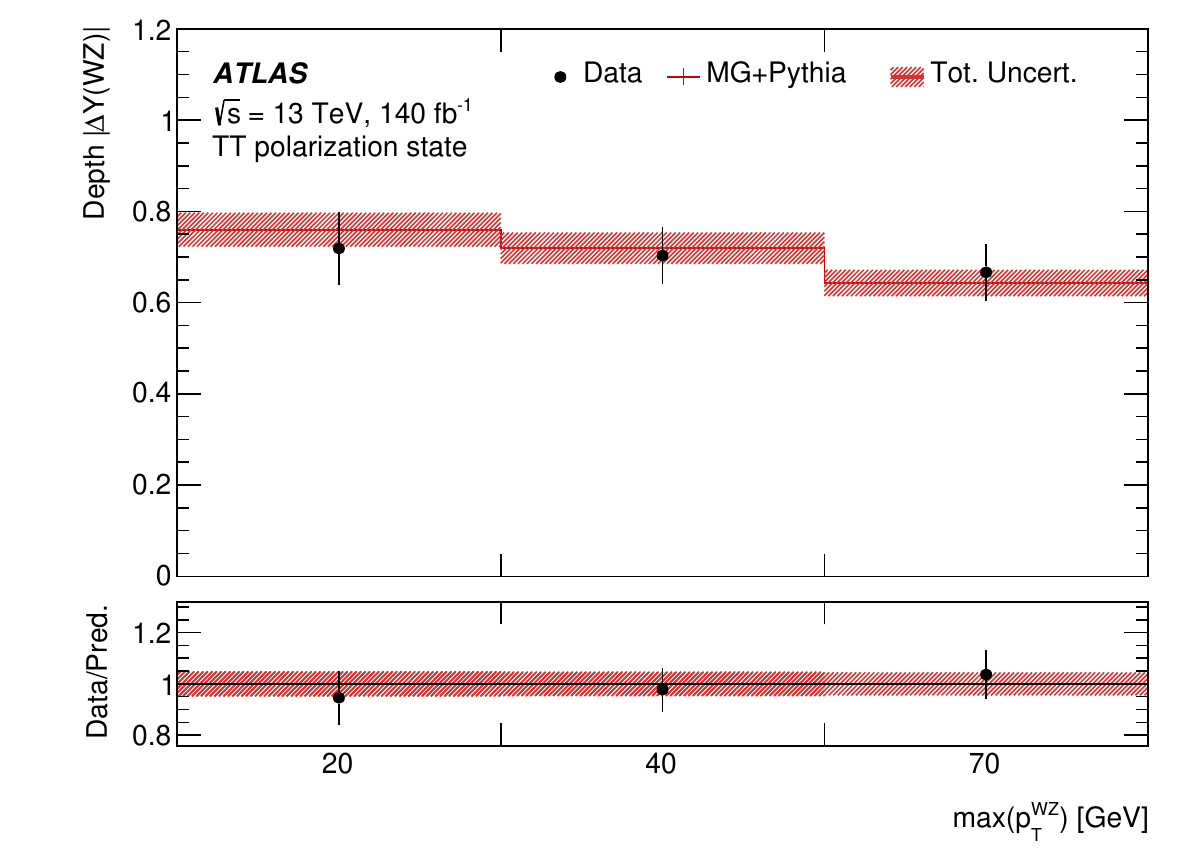} \label{fig:dip}}
\caption{
(a) Comparison between the $00+0T+T0$-subtracted normalized unfolded $|\Delta Y(WZ)|$ data distribution
and the SM prediction for $TT$ events with $p_T^{WZ}<70$ GeV; (b) The depth of the RAZ dip for the unfolded $|\Delta Y(WZ)|$ distribution of the $TT$ polarization as a function of the $p_T^{WZ}$ cut value used. The measured data are shown as black points, the vertical error bars show the size of the total uncertainty, with tick marks used to reflect the size of the statistical uncertainty only. The {\textsc{Madgraph}} 0,1j@LO prediction for $TT$ polarization state is shown in red.
}
\label{fig:RAZ}
\end{figure}

For the RAZ measurement, good agreement is observed between data and SM predictions for the $\Delta Y(WZ)$ and $\Delta Y(\ell_WZ)$ distributions in the three regions with $p_T^{WZ}<$ 20, 40, or 70 GeV. Since a dip is expected only for the $TT$ component, the estimated contributions from all background processes ($\sim 10\%$) and $WZ$ $00$, $0T$, and $T0$ polarization states ($\sim 27\%$) are subtracted from data. The $WZ$ $00$, $0T$, and $T0$ contributions are normalized to the SM predicted cross sections. The measured $|\Delta Y(\ell_W Z)|$ and $|\Delta Y(WZ)|$ distributions for the $TT$ polarization state are corrected for detector effects using an iterative Bayesian unfolding method~\cite{DAgostini:1995,Unfolding} to estimate the actual particle level normalized differential cross sections in these three $p_T^{WZ}$ regions. The unfolding procedure corrects for migrations between bins in the rapidity-difference distributions during the reconstruction of the events and applies fiducial as well as reconstruction efficiency corrections. Corrections are derived using the polarized {\textsc{Madgraph}} 0,1j@LO samples. To reduce bias due to the assumed true distribution, the method is applied iteratively, at the cost of an increased statistical uncertainty. Three iterations are used in the final unfolding procedure. Figure~\ref{fig:unf} shows the comparison between the unfolded $|\Delta Y(WZ)|$ distribution and the $TT$-only prediction for events with $p_T^{WZ}<70$ GeV. Good agreement is observed between these two distributions except the last bin where the discrepancy is 3.1 standard deviations.
In addition, the unfolding procedure is applied to inclusive events (no subtraction of contributions from $00$, $0T$ and $T0$ polarization states) in all three RAZ regions. Good agreement is also found in this case between unfolded $|\Delta Y(\ell_WZ)|$ and $|\Delta Y(WZ)|$ distributions and the SM predictions.
 
The depth of the RAZ dip, represented by the variable $\mathcal{D}$, is defined as $\mathcal{D}=1-2\times N^{\text{unf}}_{\text{central}}/N^{\text{unf}}_{\text{sides}}$, where $N^{\text{unf}}_{\text{central}}$ ($N^{\text{unf}}_{\text{sides}}$) indicates the number of events with $|\Delta Y(WZ)|<0.5$ ($0.5<|\Delta Y(\ell_WZ)|<1.5$) after the unfolding. A positive value of $\mathcal{D}$ indicates the existence of a dip.
A comparison between the measured and predicted $\mathcal{D}$ values using the $|\Delta Y(WZ)|$ distribution as a function of the $p_T^{WZ}$ cut value used is shown in Figure~\ref{fig:dip}. The same procedure is also applied for the $\Delta Y({\ell_W Z})$ distribution and again good agreement is observed between data and predictions for the depth of the dip observed.

Table~\ref{tab:yield_signal200and100} shows the expected signal and background event yields as well as the observed data in the two $00$-enhanced regions.
To measure the polarization fractions, a dedicated Boosted Decision Tree (BDT) is trained independently to separate the $00$ polarization state from the other polarization states ($0T$, $T0$, and $TT$) in each of the two signal regions defined by $p_T^Z$. The BDTs are implemented using the TMVA package~\cite{TMVA}, and the simulated {\textsc{Madgraph}} 0,1j@LO samples are used for training.
Seven variables are used for both signal regions: $\Delta Y(\ell_WZ)$, $p_T(WZ)$, the subleading lepton transverse momentum from the $Z$ boson decay, the transverse momentum of the lepton from the $W$ boson decay, \met, the cosine of the angle between the direction of the lepton from the $W$ decay in the $W$ rest frame and the direction of the $W$ boson in the $WZ$ rest frame, and the cosine of the angle between the direction of the negatively-charged lepton from the $Z$ decay in the $Z$ rest frame and the direction of the $Z$ boson in the $WZ$ rest frame.
 
\begin{table}[h!t]
\begin{center}
\small
\resizebox{0.65\linewidth}{!}{
\begin{tabular}{c |c | c}
\toprule
\textbf{Process}  & \textbf{$100<\zpT \leq 200$ GeV} & \textbf{$\zpT>200$ GeV} \\
\midrule
$W_0Z_0$            & 222 $\pm$  5  &  47.6 $\pm$ 1.5  \\
$W_0Z_T+W_TZ_0$        & 323 $\pm$ 12   &  23.7 $\pm$ 0.8   \\
$W_TZ_T$            & 856 $\pm$ 31  & 124 $\pm$ 4  \\
Prompt background   & 169 $\pm$ 18  &  24.1 $\pm$ 2.7  \\
Non-prompt background   &  68 $\pm$ 29  &  2.8  $\pm$ 1.1  \\
\midrule
Total Expected    & 1640 $\pm$ 60  & 222 $\pm$ 8 \\
\midrule
Data        & 1740    & 236   \\
\bottomrule
\end{tabular}}
\end{center}
\caption{Number of events for the expected signal, background, and data observed in the two signal regions with $p_T^{WZ}<70$ GeV and either $100<p_T^Z\leq200$ GeV or $p_T^Z>200$ GeV. Contributions from $W_0Z_0$, $W_0Z_T$, $W_TZ_0$, and $W_TZ_T$ processes are estimated from the MC simulation before the fit is performed. The uncertainties include both statistical and systematic contributions.
}
\label{tab:yield_signal200and100}
\end{table}
 
\begin{figure}[!htbp]
\centering
\subfloat[ ]{\includegraphics[width=.4\textwidth]{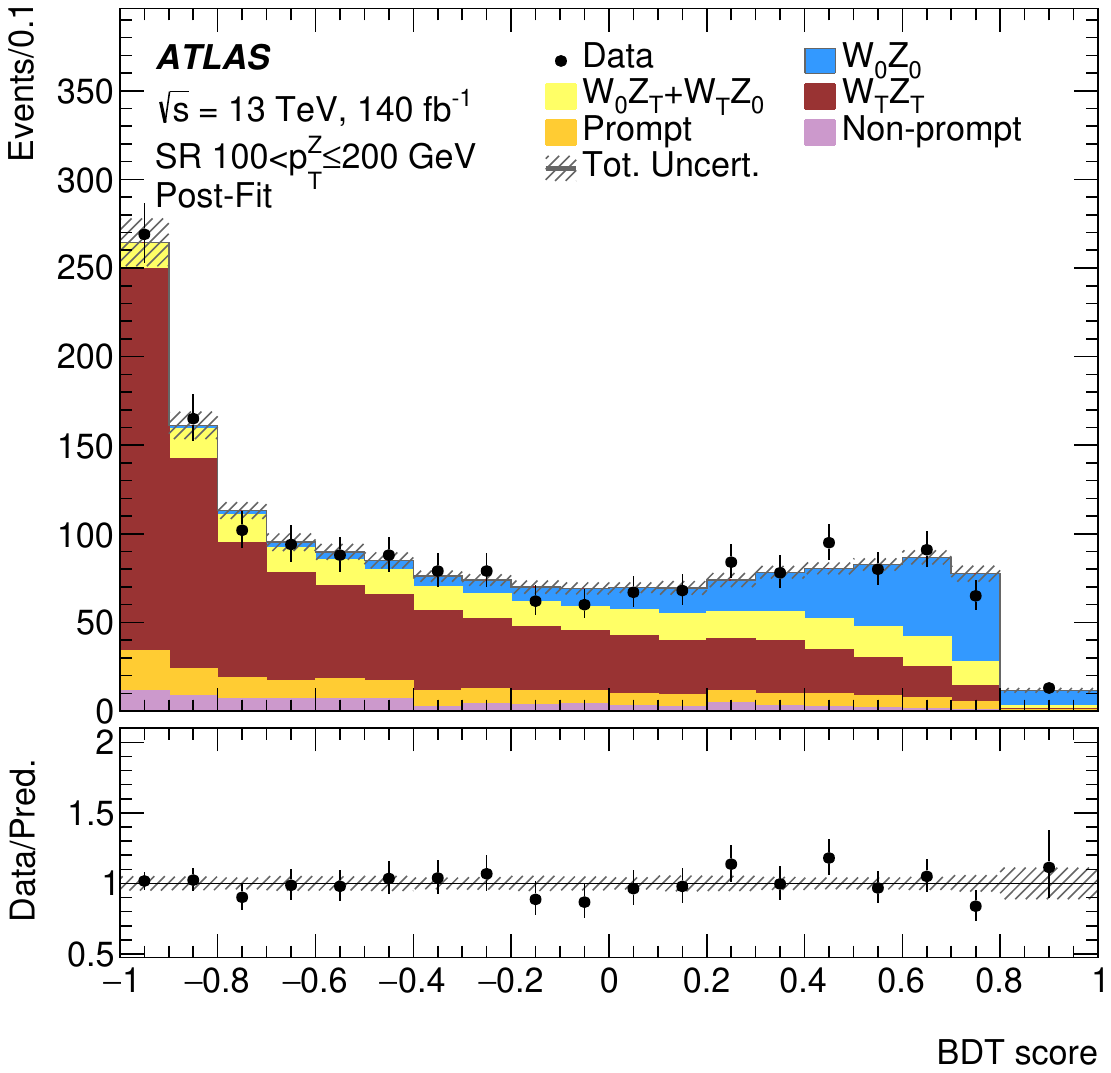}   \label{fig:100_BDT} }
\subfloat[ ]{\includegraphics[width=.4\textwidth]{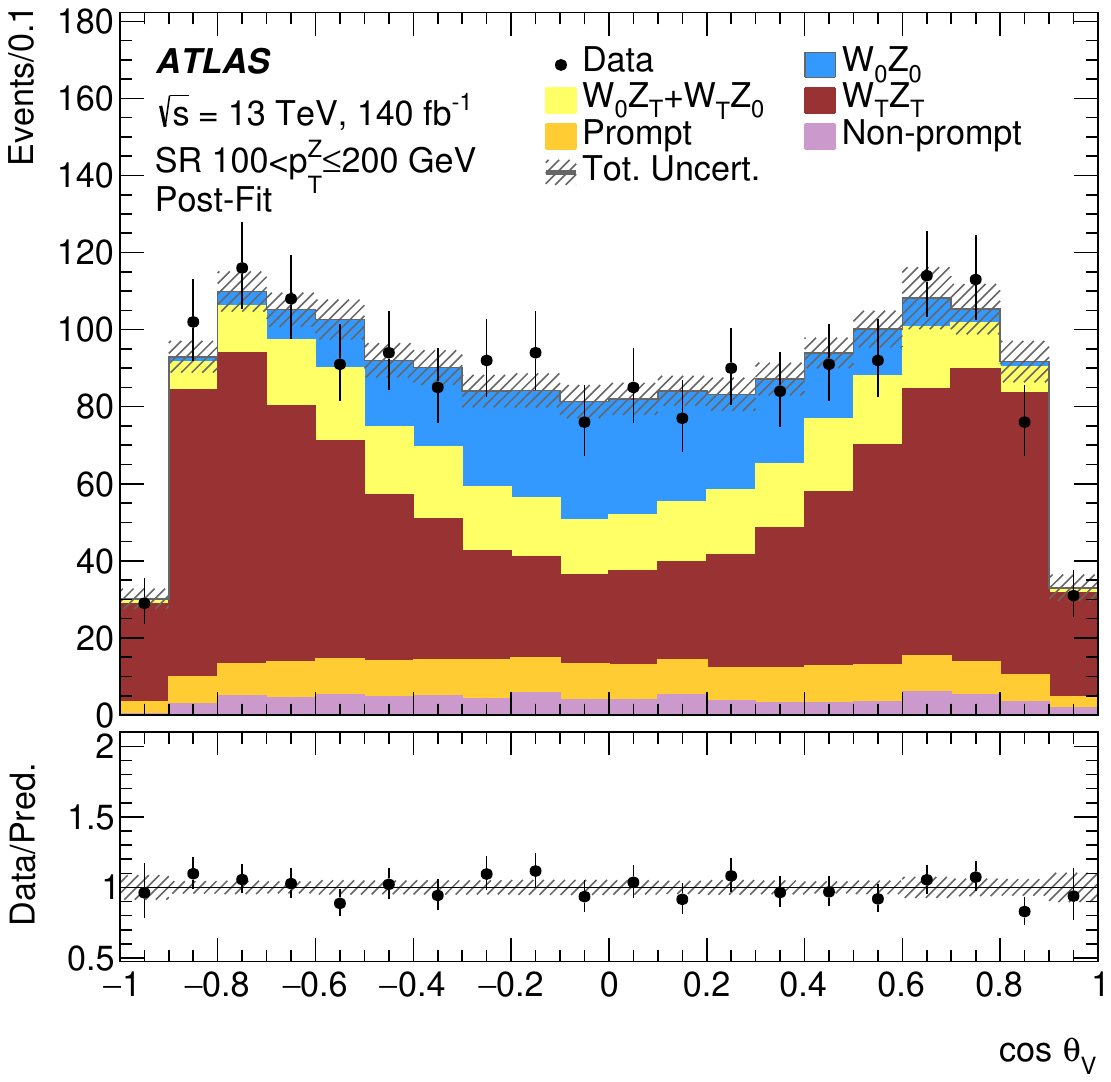}   \label{fig:100_cosThetaV}}
\caption{The comparison between the data and the post-fit SM predictions for (a) the BDT distribution and (b) the vector boson scattering angle $\cos \theta_V$ for the fiducial region with $100<p_T^Z\leq200$~GeV. The bottom panels show the ratios of the data to the post-fit SM predictions. The uncertainty bands include both the MC statistical and systematic uncertainties.}
\label{fig:BDT_fit}
\end{figure}

Binned maximum-likelihood fits~\cite{Cranmer:1456844} are performed using the BDT score distributions in the two signal regions. Each source of systematic uncertainty is implemented in the likelihood function as a nuisance parameter with a Gaussian constraint.
Three unconstrained parameters, $f_{00}$, $f_{0T+T0}$, and a signal strength modifier common to all three polarization templates, are used in the fit.
The $f_{0T}$ and $f_{T0}$ contributions are merged into a single contribution $f_{0T+T0}$ with the relative ratio according to the SM predictions, because they cannot be separated by our BDT. The fit includes the corrections for detector efficiency and acceptance effects, obtained from comparison of particle-level and reconstruction-level Monte Carlo simulations.
The BDT score distributions with background normalizations, signal normalization, and nuisance parameters adjusted by the profile-likelihood fit are shown in Figure~\ref{fig:100_BDT} for the fiducial region with $100<p_T^Z\leq200$~GeV. The comparison between the data and the post-fit predictions is also shown for $\cos \theta_V$ in Figure~\ref{fig:100_cosThetaV}.
 
The post-fit $f_{00}$ and $f_{0T+T0}$ parameters together with the expected and observed significances for $f_{00}$ are detailed in Table~\ref{table:Postfit_numbers}. For the region with $100<p_T^Z\leq200$ GeV, the observed (expected) significance is found to be 5.2 (4.3)$\sigma$. In addition, the implicit $f_{TT}$ parameter is shown as determined by \(f_{TT} = 1 - f_{00} - f_{0T+T0}\) alongside its uncertainty as determined by Gaussian error propagation.
The values of the observed correlations between \(f_{00}\) and \(f_{0T+T0}\) as determined by the fits are $-0.84$ for the region with $100<p_T^Z\leq200$ GeV and $-0.82$ for the region with $p_T^Z>200$ GeV.

The predicted fiducial polarization fractions listed in Table~\ref{table:Postfit_numbers} are calculated at the particle level in a fiducial phase space chosen to closely follow the event selection criteria using the {\textsc{Madgraph}} 0,1j@LO samples reweighted to include higher-order QCD effects and NLO electroweak corrections~\cite{ninh2023enhancing}.  Prompt leptons are dressed by adding the four-momenta of nearby prompt photons within a small cone of $\Delta R<0.1$. The distance between the two leptons from the $Z$ boson decay is required to have $\Delta R(\ell_Z^-, \ell_Z^+)>0.2$, and the distance between the negatively-charged lepton from the $Z$ boson decay and the charged lepton from the $W$ boson decay is required to have $\Delta R(\ell_Z^-, \ell_W)>0.3$. The dressed charged leptons are required to have $|\eta| < 2.5$. Leptons from the $Z~(W)$ boson decay are required to have $p_T>15 (20)$ GeV. The invariant mass of the two leptons from the $Z$ decay is required to be within $|m_{\ell\ell}-m_Z|<10$ GeV, and the transverse mass of the $W$ boson is required to be $m_T^W>30$ GeV. The $WZ$ system is required to have $p_T^{WZ}<70$ GeV, and either $100<p_T^Z\leq200$ GeV or $p_T^Z>200$ GeV.

\begin{table}[h!t]
\centering
\small
\resizebox{0.99\linewidth}{!}{
\begin{tabular}{lcc||lcc}
\toprule
& \multicolumn{2}{c||}{\textbf{Measurement}} & & \multicolumn{2}{c}{\textbf{Prediction}} \\
&  \textbf{$100<p_T^Z\leq200$ GeV} & \textbf{$p_T^Z>200$ GeV} &  &  \textbf{$100<p_T^Z\leq200$ GeV} & \textbf{$p_T^Z>200$ GeV} \\ 
\midrule
$f_{00}$     & $0.19 \pm ^{0.03}_{0.03} \text{(stat)} \pm ^{0.02}_{0.02} \text{(syst)}$ & $0.13 \pm ^{0.09}_{0.08} \text{(stat)} \pm ^{0.02}_{0.02} \text{(syst)}$ & $f_{00}$  & $0.152 \pm 0.006$ & $0.234 \pm 0.007$ \\[0.5ex]
$f_{0T+T0}$  & $0.18 \pm ^{0.07}_{0.08} \text{(stat)} \pm ^{0.05}_{0.06} \text{(syst)}$ & $0.23 \pm ^{0.17}_{0.18} \text{(stat)} \pm ^{0.06}_{0.10} \text{(syst)}$ & $f_{0T}$ & $0.120 \pm 0.002$ & $0.062 \pm 0.002$ \\[0.5ex]
$f_{TT}$ & $0.63 \pm ^{0.05}_{0.05} \text{(stat)} \pm ^{0.04}_{0.04} \text{(syst)}$ & $0.64 \pm ^{0.12}_{0.12} \text{(stat)} \pm ^{0.06}_{0.06} \text{(syst)}$ & $f_{T0}$ & $0.109 \pm 0.001$ & $0.058 \pm 0.001$ \\[0.5ex]
$f_{00}$ obs (exp) sig.        & 5.2 (4.3) $\sigma$ & 1.6 (2.5) $\sigma$ & $f_{TT}$  & $0.619 \pm 0.007$ & $0.646 \pm 0.008$  \\
\bottomrule
\end{tabular}
}
\caption{Measured diboson polarization fractions in the two signal regions with $p_T^{WZ}<70$ GeV and $100<p_T^Z\leq200$ GeV or $p_T^Z>200$ GeV using three unconstrained parameters. The SM predicted fractions for all four polarization states are also shown.}
\label{table:Postfit_numbers}
\end{table}
 
Another fit is performed using two unconstrained parameters, $f_{00}$ and the total number of $WZ$ events. The mixed $f_{0T+T0}$ and doubly-transversal $f_{TT}$ contributions are combined into a single contribution ($f_{XX}$) defined as $f_{XX}=1-f_{00}$. The $00$ fraction is found to be $f_{00}=0.17 \pm ^{0.02}_{0.02} \text{(stat)} \pm ^{0.01}_{0.02} \text{(syst)}$ for the region with $100<p_T^Z\leq200$~GeV and $0.16 \pm ^{0.05}_{0.05} \text{(stat)} \pm ^{0.02}_{0.03} \text{(syst)}$ for the region with $p_T^Z>200$~GeV. The corresponding observed (expected) significance for a non-zero $f_{00}$ value is 7.7 (6.9) $\sigma$ for the first region and 3.2 (4.2) $\sigma$ for the second region.
This fit results in better sensitivities for $f_{00}$, however, it may be less conservative as the ratio of $f_{0T+T0}$ and $f_{TT}$ is assumed to have the value predicted by the SM.
 
In summary, studies of the RAZ effect and the energy dependence of diboson polarization fractions in $WZ$ production are presented in this Letter.
The measurements use leptonic decay modes of the gauge bosons to electrons or muons.
Significant dips are observed in the $\Delta Y(\ell_WZ)$ and $\Delta Y(WZ)$ distributions for inclusive $TT$ events with different $p_T^{WZ}$ cuts applied, indicating the presence of the RAZ effect in $WZ$ production at the LHC. Unfolded $|\Delta Y(\ell_WZ)|$ and $|\Delta Y(WZ)|$ distributions are also measured and compared to theoretical predictions.
Diboson polarization fractions are measured in two signal regions with enhanced longitudinal polarization for both bosons. The measured fractions are found to be consistent with the SM predictions.
A non-zero fraction of events where both bosons are longitudinally polarized is measured with an observed significance of 5.2 $\sigma$ (1.6 $\sigma$) in the phase space with $100<p_T^Z\leq200$ GeV ($p_T^Z>200$~GeV).
 
\section*{Acknowledgements}


We thank CERN for the very successful operation of the LHC and its injectors, as well as the support staff at
CERN and at our institutions worldwide without whom ATLAS could not be operated efficiently.
 
The crucial computing support from all WLCG partners is acknowledged gratefully, in particular from CERN, the ATLAS Tier-1 facilities at TRIUMF/SFU (Canada), NDGF (Denmark, Norway, Sweden), CC-IN2P3 (France), KIT/GridKA (Germany), INFN-CNAF (Italy), NL-T1 (Netherlands), PIC (Spain), RAL (UK) and BNL (USA), the Tier-2 facilities worldwide and large non-WLCG resource providers. Major contributors of computing resources are listed in Ref.~\cite{ATL-SOFT-PUB-2023-001}.
 
We gratefully acknowledge the support of ANPCyT, Argentina; YerPhI, Armenia; ARC, Australia; BMWFW and FWF, Austria; ANAS, Azerbaijan; CNPq and FAPESP, Brazil; NSERC, NRC and CFI, Canada; CERN; ANID, Chile; CAS, MOST and NSFC, China; Minciencias, Colombia; MEYS CR, Czech Republic; DNRF and DNSRC, Denmark; IN2P3-CNRS and CEA-DRF/IRFU, France; SRNSFG, Georgia; BMBF, HGF and MPG, Germany; GSRI, Greece; RGC and Hong Kong SAR, China; ISF and Benoziyo Center, Israel; INFN, Italy; MEXT and JSPS, Japan; CNRST, Morocco; NWO, Netherlands; RCN, Norway; MNiSW, Poland; FCT, Portugal; MNE/IFA, Romania; MESTD, Serbia; MSSR, Slovakia; ARRS and MIZ\v{S}, Slovenia; DSI/NRF, South Africa; MICINN, Spain; SRC and Wallenberg Foundation, Sweden; SERI, SNSF and Cantons of Bern and Geneva, Switzerland; MOST, Taipei; TENMAK, T\"urkiye; STFC, United Kingdom; DOE and NSF, United States of America.
 
Individual groups and members have received support from BCKDF, CANARIE, CRC and DRAC, Canada; CERN-CZ, PRIMUS 21/SCI/017 and UNCE SCI/013, Czech Republic; COST, ERC, ERDF, Horizon 2020, ICSC-NextGenerationEU and Marie Sk{\l}odowska-Curie Actions, European Union; Investissements d'Avenir Labex, Investissements d'Avenir Idex and ANR, France; DFG and AvH Foundation, Germany; Herakleitos, Thales and Aristeia programmes co-financed by EU-ESF and the Greek NSRF, Greece; BSF-NSF and MINERVA, Israel; Norwegian Financial Mechanism 2014-2021, Norway; NCN and NAWA, Poland; La Caixa Banking Foundation, CERCA Programme Generalitat de Catalunya and PROMETEO and GenT Programmes Generalitat Valenciana, Spain; G\"{o}ran Gustafssons Stiftelse, Sweden; The Royal Society and Leverhulme Trust, United Kingdom.
 
In addition, individual members wish to acknowledge support from CERN: European Organization for Nuclear Research (CERN PJAS); Chile: Agencia Nacional de Investigaci\'on y Desarrollo (FONDECYT 1190886, FONDECYT 1210400, FONDECYT 1230812, FONDECYT 1230987); China: Chinese Ministry of Science and Technology (MOST-2023YFA1605700), National Natural Science Foundation of China (NSFC - 12175119, NSFC 12275265, NSFC-12075060); Czech Republic: PRIMUS Research Programme (PRIMUS/21/SCI/017); EU: H2020 European Research Council (ERC - 101002463); European Union: European Research Council (ERC - 948254, ERC 101089007), Horizon 2020 Framework Programme (MUCCA - CHIST-ERA-19-XAI-00), European Union, Future Artificial Intelligence Research (FAIR-NextGenerationEU PE00000013), Italian Center for High Performance Computing, Big Data and Quantum Computing (ICSC, NextGenerationEU); France: Agence Nationale de la Recherche (ANR-20-CE31-0013, ANR-21-CE31-0013, ANR-21-CE31-0022), Investissements d'Avenir Labex (ANR-11-LABX-0012); Germany: Baden-Württemberg Stiftung (BW Stiftung-Postdoc Eliteprogramme), Deutsche Forschungsgemeinschaft (DFG - 469666862, DFG - CR 312/5-2); Italy: Istituto Nazionale di Fisica Nucleare (ICSC, NextGenerationEU); Japan: Japan Society for the Promotion of Science (JSPS KAKENHI JP21H05085, JSPS KAKENHI JP22H01227, JSPS KAKENHI JP22H04944, JSPS KAKENHI JP22KK0227); Netherlands: Netherlands Organisation for Scientific Research (NWO Veni 2020 - VI.Veni.202.179); Norway: Research Council of Norway (RCN-314472); Poland: Polish National Agency for Academic Exchange (PPN/PPO/2020/1/00002/U/00001), Polish National Science Centre (NCN 2021/42/E/ST2/00350, NCN OPUS nr 2022/47/B/ST2/03059, NCN UMO-2019/34/E/ST2/00393, UMO-2020/37/B/ST2/01043, UMO-2021/40/C/ST2/00187, UMO-2022/47/O/ST2/00148); Slovenia: Slovenian Research Agency (ARIS grant J1-3010); Spain: BBVA Foundation (LEO22-1-603), Generalitat Valenciana (Artemisa, FEDER, IDIFEDER/2018/048), Ministry of Science and Innovation (MCIN \& NextGenEU PCI2022-135018-2, MICIN \& FEDER PID2021-125273NB, RYC2019-028510-I, RYC2020-030254-I, RYC2021-031273-I, RYC2022-038164-I), PROMETEO and GenT Programmes Generalitat Valenciana (CIDEGENT/2019/023, CIDEGENT/2019/027); Sweden: Swedish Research Council (VR 2018-00482, VR 2022-03845, VR 2022-04683, VR grant 2021-03651), Knut and Alice Wallenberg Foundation (KAW 2017.0100, KAW 2018.0157, KAW 2018.0458, KAW 2019.0447, KAW 2022.0358); Switzerland: Swiss National Science Foundation (SNSF - PCEFP2\_194658); United Kingdom: Leverhulme Trust (Leverhulme Trust RPG-2020-004), Royal Society (NIF-R1-231091); United States of America: U.S. Department of Energy (ECA DE-AC02-76SF00515), Neubauer Family Foundation.


We thank Duc Ninh Le, Da Liu, and Giovanni Pelliccioli for providing calculations with higher-order QCD and electroweak corrections in $WZ$ production with different polarizations.
 
\appendix
\part*{Appendix}
\begin{figure}[htbp]
\centering
\subfloat[ ]{\includegraphics[width=0.45\textwidth]{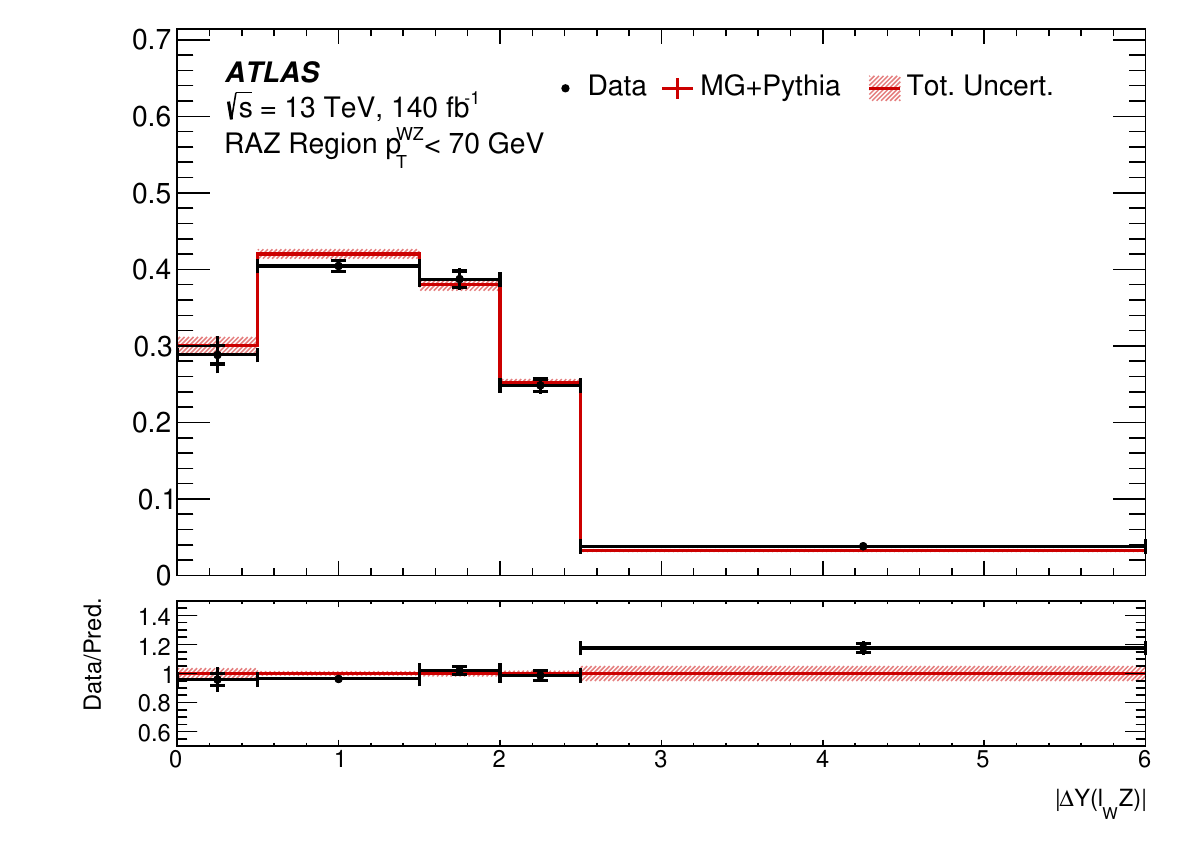}\label{fig:RAZ_PRL0_a}}
\subfloat[ ]{\includegraphics[width=0.45\textwidth]{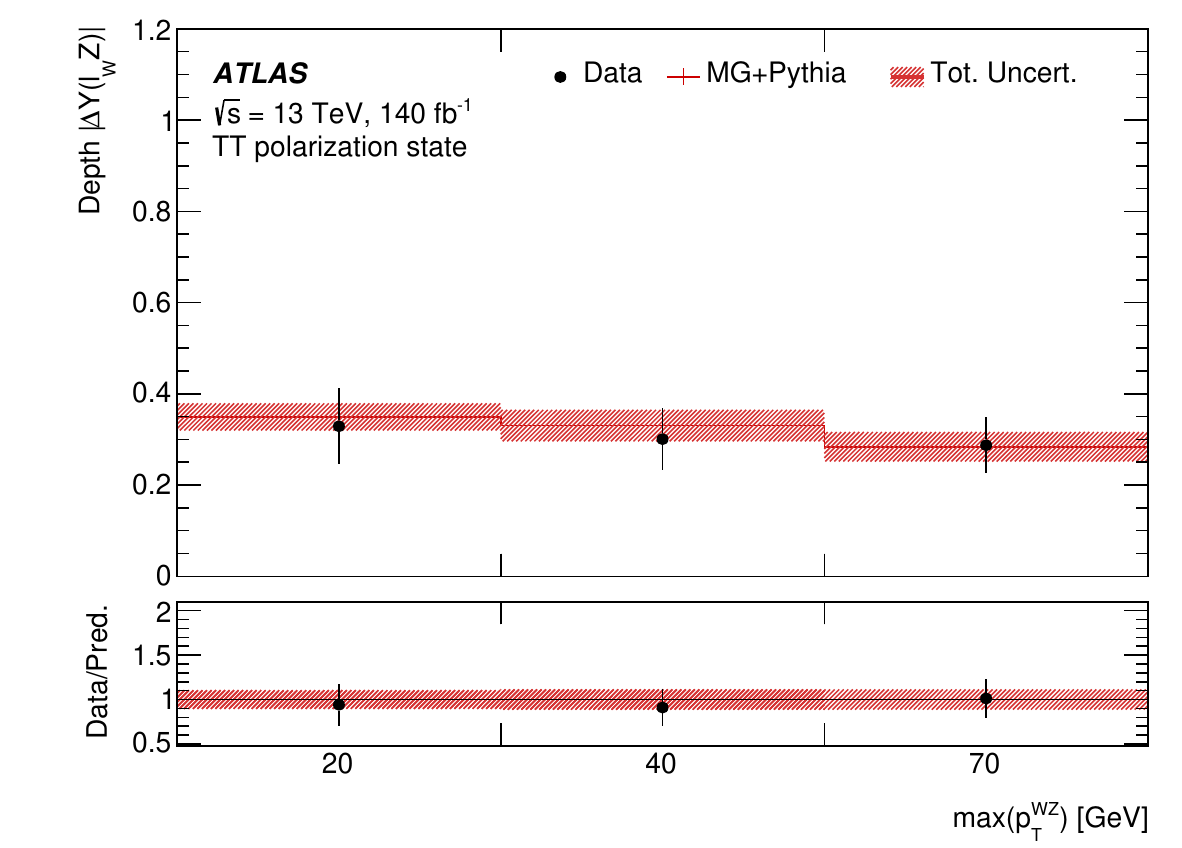} \label{fig:RAZ_PRL0_b}}
\caption{
(a) Comparison between the $00+0T+T0$-subtracted normalized unfolded $|\Delta Y(\ell_WZ)|$ data distribution
and the SM prediction for $TT$ events with $p_T^{WZ}<70$ GeV; (b) The depth of the RAZ dip for the unfolded $|\Delta Y(\ell_WZ)|$ distribution of the $TT$ polarization as a function of the $p_T^{WZ}$ cut value used. The measured data are shown as black points, the vertical error bars show the size of the total uncertainty, with tick marks used to reflect the size of the statistical uncertainty only. The {\textsc{Madgraph}} 0,1j@LO prediction for $TT$ polarization state is shown in red.}
\label{fig:RAZ_appendix_PRL0}
\end{figure}

\addcontentsline{toc}{part}{Appendix}
The unfolded distribution of the $\Delta Y({\ell_W Z})$ considering as signal the $TT$ component for events with $p_T^{WZ}<70$ GeV is shown in Figure~\ref{fig:RAZ_PRL0_a}. A comparison between the measured and predicted $\mathcal{D}$ values using the $|\Delta Y(\ell_WZ)|$ distribution as a function of the $p_T^{WZ}$ cut value used is shown in Figure~\ref{fig:RAZ_PRL0_b}. Good agreement is observed between data and predictions for the unfolded distributions and the depth of the dip.
 
\begin{figure}[hbt!]
\centering
\subfloat[ ]{\includegraphics[width=.42\textwidth]{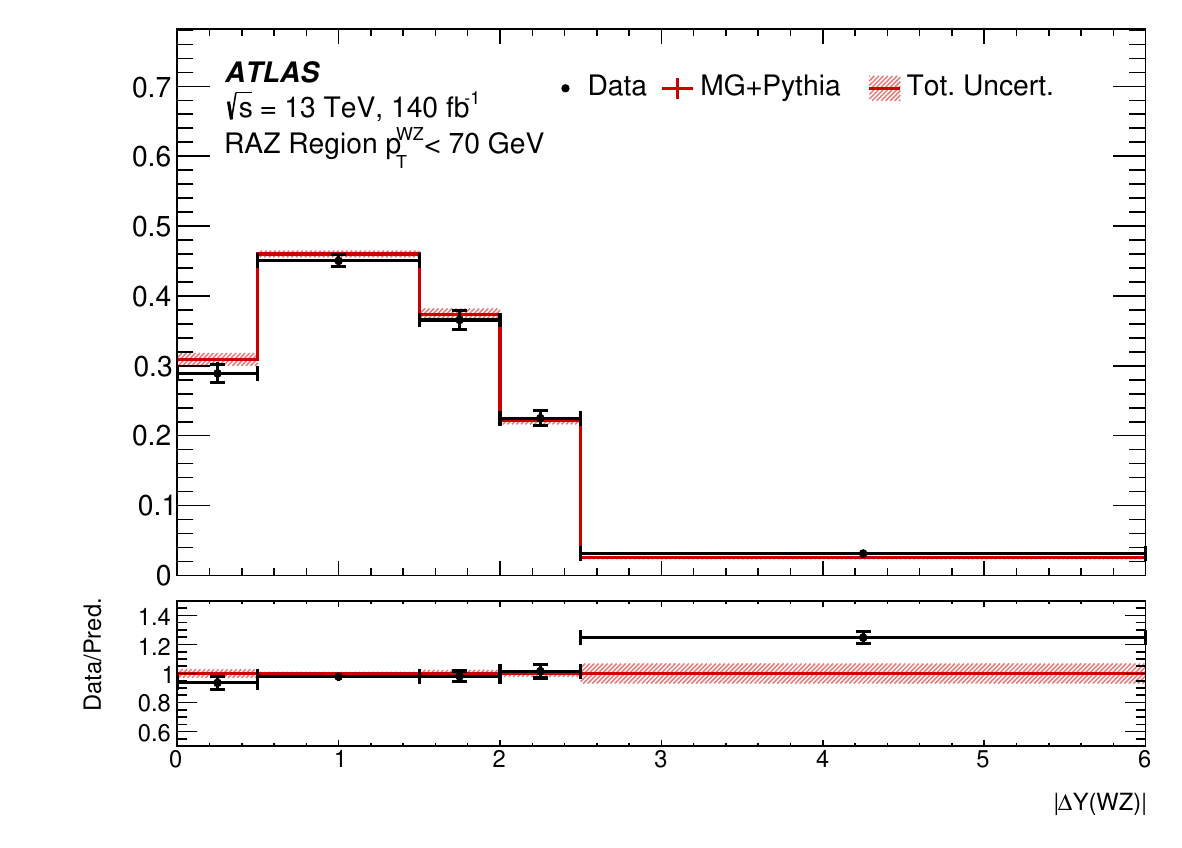} \label{fig:RAZ_PRL_a}}
\subfloat[ ]{\includegraphics[width=.42\textwidth]{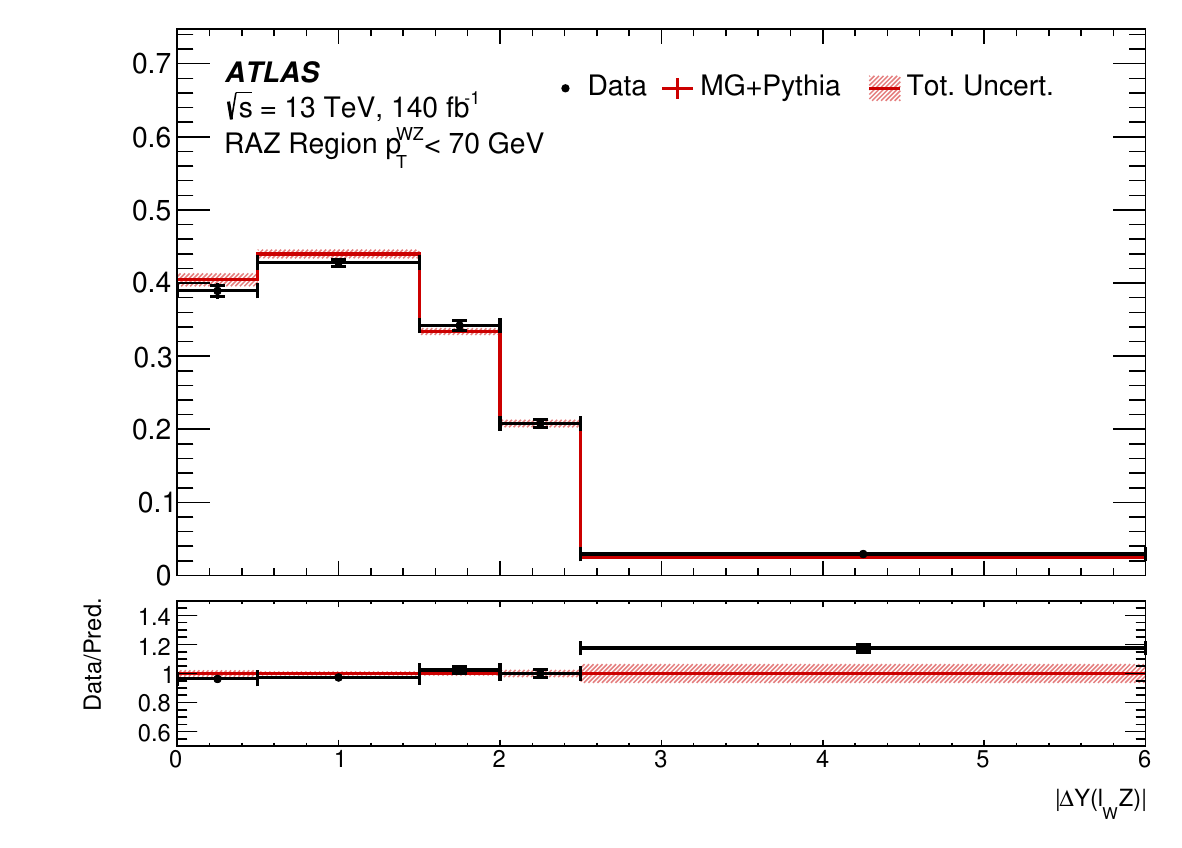} \label{fig:RAZ_PRL_b}}
\caption{
Comparison between the unfolded (a) $|\Delta Y(WZ)|$ and (b) $|\Delta Y(\ell_WZ)|$ distributions and the SM prediction for inclusive $WZ$ events with $p_T^{WZ}<70$ GeV. The measured data are shown as black points with horizontal bars indicating the bin range and the vertical inner (outer) error bars representing the statistical (total) uncertainty. The {\textsc{Madgraph}} 0,1j@LO prediction for the sum of all polarization states is shown in red.}
\label{fig:RAZ_appendix_PRL}
\end{figure}

Unfolded $|\Delta Y(WZ)|$ and $|\Delta Y(\ell_WZ)|$ distributions for inclusive $WZ$ events with $p_T^{WZ}<20$, 40, or 70 GeV are also measured and compared to theoretical predictions. This alternative signal definition considers all polarization states together and avoids assumptions on the $00$, $0T$, and $T0$ cross sections. Figure~\ref{fig:RAZ_appendix_PRL} shows the comparison plots for events with $p_T^{WZ}<70$ GeV. The uncertainties on the unfolded distributions are dominated by the data statistical uncertainty in all bins. In general, the SM predictions agree well with the measured normalized differential cross sections within the quoted uncertainties. The largest difference is observed in the last bin of the $|\Delta Y(WZ)|$ distribution reaching a local significance of 2.6 standard deviations.
 
\clearpage

\printbibliography

@article{Grazzini:2015hta,
    author = "Grazzini, Massimiliano and Kallweit, Stefan and Rathlev, Dirk",
    title = "{$ZZ$ production at the LHC: fiducial cross sections and distributions in NNLO QCD}",
    eprint = "1507.06257",
    archivePrefix = "arXiv",
    primaryClass = "hep-ph",
    doi = "10.1016/j.physletb.2015.09.055",
    journal = "Phys. Lett. B",
    volume = "750",
    pages = "407--410",
    year = "2015"
}

@article{Binoth:2008kt,
    author = "Binoth, T. and Ossola, G. and Papadopoulos, C. G. and Pittau, R.",
    title = "{NLO QCD corrections to tri-boson production}",
    eprint = "0804.0350",
    archivePrefix = "arXiv",
    primaryClass = "hep-ph",
    doi = "10.1088/1126-6708/2008/06/082",
    journal = "JHEP",
    volume = "06",
    pages = "082",
    year = "2008"
}

@article{Frederix:2021agh,
    author = "Frederix, Rikkert and Tsinikos, Ioannis",
    title = "{On improving NLO merging for $ \mathrm{t}\overline{\mathrm{t}}\mathrm{W} $ production}",
    eprint = "2108.07826",
    archivePrefix = "arXiv",
    primaryClass = "hep-ph",
    reportNumber = "LU-TP 21-34",
    doi = "10.1007/JHEP11(2021)029",
    journal = "JHEP",
    volume = "11",
    pages = "029",
    year = "2021"
}

@article{Denner:2023eti,
    author = "Denner, Ansgar and Lombardi, Daniele and Pelliccioli, Giovanni",
    title = "{Complete NLO corrections to off-shell $ \textrm{t}\overline{\textrm{t}}\textrm{Z} $ production at the LHC}",
    eprint = "2306.13535",
    archivePrefix = "arXiv",
    primaryClass = "hep-ph",
    reportNumber = "MPP-2023-132",
    doi = "10.1007/JHEP09(2023)072",
    journal = "JHEP",
    volume = "09",
    pages = "072",
    year = "2023"
}

@article{Capdevilla:2019zbx,
    author = "Capdevilla, Rodolfo and Harnik, Roni and Martin, Adam",
    title = "{The radiation valley and exotic resonances in $W\gamma$ production at the LHC}",
    eprint = "1912.08234",
    archivePrefix = "arXiv",
    primaryClass = "hep-ph",
    reportNumber = "FERMILAB-PUB-19-630-T",
    doi = "10.1007/JHEP03(2020)117",
    journal = "JHEP",
    volume = "03",
    pages = "117",
    year = "2020"
}

@article{Dixon:1999di,
    author = "Dixon, Lance and Kunszt, Z. and Signer, A.",
    title = {Vector boson pair production in hadronic collisions at $O({\ensuremath{\alpha}}_{s}):$ Lepton correlations and anomalous couplings},
    eprint = "hep-ph/9907305",
    archivePrefix = "arXiv",
    reportNumber = "SLAC-PUB-8188, ETH-TH-99-10, DTP-99-72",
    doi = "10.1103/PhysRevD.60.114037",
    journal = "Phys. Rev. D",
    volume = "60",
    pages = "114037",
    year = "1999"
}

@article{Campbell:2011bn,
    author = "Campbell, John M. and Ellis, R. Keith and Williams, Ciaran",
    title = "{Vector boson pair production at the LHC}",
    eprint = "1105.0020",
    archivePrefix = "arXiv",
    primaryClass = "hep-ph",
    reportNumber = "FERMILAB-PUB-11-182-T",
    doi = "10.1007/JHEP07(2011)018",
    journal = "JHEP",
    volume = "07",
    pages = "018",
    year = "2011"
}

@article{Sherpa:2019gpd,
    author = "Bothmann, Enrico and others",
    collaboration = "Sherpa",
    title = "{Event Generation with Sherpa 2.2}",
    eprint = "1905.09127",
    archivePrefix = "arXiv",
    primaryClass = "hep-ph",
    reportNumber = "FERMILAB-PUB-19-218-T, SLAC-PUB-17433, IPPP/19/42, MCNET-19-11",
    doi = "10.21468/SciPostPhys.7.3.034",
    journal = "SciPost Phys.",
    volume = "7",
    number = "3",
    pages = "034",
    year = "2019"
}

@article{Avoni:2018iuv,
    author = "Avoni, G. and others",
    title = "{The new LUCID-2 detector for luminosity measurement and monitoring in ATLAS}",
    doi = "10.1088/1748-0221/13/07/P07017",
    journal = "JINST",
    volume = "13",
    number = "07",
    pages = "P07017",
    year = "2018"
}

@Booklet{Unfolding,
      author         = "D'Agostini, G.",
      title          = "{Improved iterative Bayesian unfolding}",
      year           = "2010",
      eprint         = "1010.0632",
      archivePrefix  = "arXiv",
      primaryClass   = "physics.data-an",
      doi            = "10.48550/arXiv.1010.0632",
}

@Booklet{Cranmer:1456844,
      xauthor        = "{ROOT Collaboration}",
      author       = "Cranmer, Kyle and Lewis, George and Moneta, Lorenzo and
                       Shibata, Akira and Verkerke, Wouter",
      xcollaboration= "ROOT",
      title         = "{HistFactory: A tool for creating statistical models for
                       use with RooFit and RooStats}",
      institution   = "CERN",
      reportNumber  = "CERN-OPEN-2012-016",
      address       = "New York",
      year          = "2012",
      howpublished   = "{CERN-OPEN-2012-016}",
      url           = "https://cds.cern.ch/record/1456844",
      doi           = "10.17181/CERN-OPEN-2012-016",
}

@article{Baur:1994ia,
    author = "Baur, U. and Han, Tao and Ohnemus, J.",
    title = "{Amplitude zeros in $W^\pm Z$ production}",
    eprint = "hep-ph/9403248",
    archivePrefix = "arXiv",
    reportNumber = "FSU-HEP-940307, UCD-94-5",
    doi = "10.1103/PhysRevLett.72.3941",
    journal = "Phys. Rev. Lett.",
    volume = "72",
    pages = "3941--3944",
    year = "1994"
}

@article{Denner:2020eck,
    author = "Denner, Ansgar and Pelliccioli, Giovanni",
    title = "{NLO QCD predictions for doubly-polarized $WZ$ production at the LHC}",
    eprint = "2010.07149",
    archivePrefix = "arXiv",
    primaryClass = "hep-ph",
    doi = "10.1016/j.physletb.2021.136107",
    journal = "Phys. Lett. B",
    volume = "814",
    pages = "136107",
    year = "2021"
}

@article{Pelliccioli:2023zpd,
    author = "Pelliccioli, Giovanni and Zanderighi, Giulia",
    title = "{Polarised-boson pairs at the LHC with NLOPS accuracy}",
    eprint = "2311.05220",
    archivePrefix = "arXiv",
    primaryClass = "hep-ph",
    doi = "10.1140/epjc/s10052-023-12347-4",
    journal = "Eur. Phys. J. C",
    volume = "84",
    number = "1",
    pages = "16",
    year = "2024"
}

@article{Baur:1994aj,
      author         = "Baur, U. and Han, Tao and Ohnemus, J.",
      title          = "{$W Z$ production at hadron colliders: Effects of
                        nonstandard $W W Z$ couplings and QCD corrections}",
      journal        = "Phys. Rev. D",
      volume         = "51",
      pages          = "3381-3407",
      doi            = "10.1103/PhysRevD.51.3381",
      year           = "1995",
      eprint         = "hep-ph/9410266",
      archivePrefix  = "arXiv",
      primaryClass   = "hep-ph",
      reportNumber   = "FSU-HEP-941010, UCD-94-22",
      SLACcitation   = "%%CITATION = HEP-PH/9410266;%%",
}

@article{Frye:2015rba,
    author = "Frye, Christopher and Freytsis, Marat and Scholtz, Jakub and Strassler, Matthew J.",
    title = "{Precision Diboson Observables for the LHC}",
    eprint = "1510.08451",
    archivePrefix = "arXiv",
    primaryClass = "hep-ph",
    doi = "10.1007/JHEP03(2016)171",
    journal = "JHEP",
    volume = "03",
    pages = "171",
    year = "2016"
}

@article{Frixione:1992pj,
    author = "Frixione, S. and Nason, P. and Ridolfi, G.",
    title = "{Strong corrections to $WZ$ production at hadron colliders}",
    reportNumber = "GEF-TH-2-1992, UPRF-92-323",
    doi = "10.1016/0550-3213(92)90668-2",
    journal = "Nucl. Phys. B",
    volume = "383",
    pages = "3--44",
    year = "1992"
}

@article{Bilchak:1984gv,
    author = "Bilchak, C. L. and Brown, R. W. and Stroughair, J. D.",
    title = "{$W^\pm$ and $Z^0$ polarization in pair production: Dominant helicities}",
    doi = "10.1103/PhysRevD.29.375",
    journal = "Phys. Rev. D",
    volume = "29",
    pages = "375--386",
    year = "1984"
}

@article{D0:2008abl,
    author = "{D0 Collaboration}",
    xauthor = "Abazov, V. M. and others",
    xcollaboration = "D0",
    title = "{First study of the radiation-amplitude zero in $W \gamma$ production and limits on anomalous $W W \gamma$ couplings at $\sqrt{s} = 1.96$ TeV}",
    eprint = "0803.0030",
    archivePrefix = "arXiv",
    primaryClass = "hep-ex",
    reportNumber = "FERMILAB-PUB-08-052-E",
    doi = "10.1103/PhysRevLett.100.241805",
    journal = "Phys. Rev. Lett.",
    volume = "100",
    pages = "241805",
    year = "2008"
}

@article{Franceschini:2017xkh,
      author         = "Franceschini, Roberto and Panico, Giuliano and Pomarol,
                        Alex and Riva, Francesco and Wulzer, Andrea",
      title          = "{Electroweak Precision Tests in High-Energy Diboson
                        Processes}",
      journal        = "JHEP",
      volume         = "02",
      year           = "2018",
      pages          = "111",
      doi            = "10.1007/JHEP02(2018)111",
      eprint         = "1712.01310",
      archivePrefix  = "arXiv",
      primaryClass   = "hep-ph",
      reportNumber   = "CERN-TH-2017-252, RM3-TH-17-1",
      SLACcitation   = "%%CITATION = ARXIV:1712.01310;%%"
}

@article{Liu:2019vid,
      author         = "Liu, Da and Wang, Lian-Tao",
      title          = "{Prospects for precision measurement of diboson processes
                        in the semileptonic decay channel in future LHC runs}",
      journal        = "Phys. Rev. D",
      volume         = "99",
      year           = "2019",
      number         = "5",
      pages          = "055001",
      doi            = "10.1103/PhysRevD.99.055001",
      eprint         = "1804.08688",
      archivePrefix  = "arXiv",
      primaryClass   = "hep-ph",
      SLACcitation   = "%%CITATION = ARXIV:1804.08688;%%"
}

@article{Huang:2020iya,
    author = "Huang, Li and Lane, Samuel D. and Lewis, Ian M. and Liu, Zhen",
    title = "{Electroweak Restoration at the LHC and Beyond: The $Vh$ Channel}",
    eprint = "2012.00774",
    archivePrefix = "arXiv",
    primaryClass = "hep-ph",
    doi = "10.1103/PhysRevD.103.053007",
    journal = "Phys. Rev. D",
    volume = "103",
    number = "5",
    pages = "053007",
    year = "2021"
}

@article{Ohnemus:1991gb,
  title = {Order-${\mathrm{\ensuremath{\alpha}}}_{\mathit{s}}$ calculation of hadronic ${\mathit{W}}^{\ifmmode\pm\else\textpm\fi{}}$Z production},
  author = {Ohnemus, J.},
  journal = {Phys. Rev. D},
  volume = {44},
  pages = {3477--3489},
  numpages = {0},
  year = {1991},
  month = Dec,
  publisher = {American Physical Society},
  doi = {10.1103/PhysRevD.44.3477},
}

@article{Bierweiler:2013dja,
    author = {Bierweiler, Anastasiya and Kasprzik, Tobias and K\"uhn, Johann H.},
    title = "{Vector-boson pair production at the LHC to $\mathcal{O}(\alpha^3)$ accuracy}",
    eprint = "1305.5402",
    archivePrefix = "arXiv",
    primaryClass = "hep-ph",
    doi = "10.1007/JHEP12(2013)071",
    journal = "JHEP",
    volume = "12",
    pages = "071",
    year = "2013"
}

@article{Baglio:2013toa,
  title = {Erratum: Massive gauge boson pair production at the LHC: A next-to-leading order story [Phys. Rev. D 88, 113005 (2013)]},
  author = {Baglio, Julien and Ninh, Le Duc and Weber, Marcus M.},
  journal = {Phys. Rev. D},
  volume = {94},
  pages = {099902},
  numpages = {1},
  year = {2016},
  month = Nov,
  publisher = {American Physical Society},
  doi = {10.1103/PhysRevD.94.099902},
  eprint = "1307.4331",
  archivePrefix = "arXiv",
  primaryClass = "hep-ph",
}

@article{Biedermann:2017oae,
    author = "Biedermann, Benedikt and Denner, Ansgar and Hofer, Lars",
    title = "{Next-to-leading-order electroweak corrections to the production of three charged leptons plus missing energy at the LHC}",
    eprint = "1708.06938",
    archivePrefix = "arXiv",
    primaryClass = "hep-ph",
    reportNumber = "ICCUB-17-016",
    doi = "10.1007/JHEP10(2017)043",
    journal = "JHEP",
    volume = "10",
    pages = "043",
    year = "2017"
}

@article{Grazzini:2016swo,
    author = "Grazzini, Massimiliano and Kallweit, Stefan and Rathlev, Dirk and Wiesemann, Marius",
    title = "{$W^{\pm}Z$ production at hadron colliders in NNLO QCD}",
    eprint = "1604.08576",
    archivePrefix = "arXiv",
    primaryClass = "hep-ph",
    reportNumber = "ZU-TH-15-16, MITP-16-037, NSF-KITP-16-046, DESY-16-074",
    doi = "10.1016/j.physletb.2016.08.017",
    journal = "Phys. Lett. B",
    volume = "761",
    pages = "179--183",
    year = "2016"
}

@article{Grazzini:2017ckn,
    author = "Grazzini, Massimiliano and Kallweit, Stefan and Rathlev, Dirk and Wiesemann, Marius",
    title = "{$W^\pm Z$ production at the LHC: fiducial cross sections and distributions in NNLO QCD}",
    eprint = "1703.09065",
    archivePrefix = "arXiv",
    primaryClass = "hep-ph",
    reportNumber = "ZU-TH-06-17, CERN-TH-2017-065",
    doi = "10.1007/JHEP05(2017)139",
    journal = "JHEP",
    volume = "05",
    pages = "139",
    year = "2017"
}

@article{Grazzini:2019jkl,
    author = "Grazzini, Massimiliano and Kallweit, S. and Lindert, Jonas M. and Pozzorini, Stefano and Wiesemann, Marius",
    title = "{NNLO QCD + NLO EW with Matrix+OpenLoops: precise predictions for vector-boson pair production}",
    eprint = "1912.00068",
    archivePrefix = "arXiv",
    primaryClass = "hep-ph",
    reportNumber = "MPP-2019-175, ZU-TH 49/19",
    doi = "10.1007/JHEP02(2020)087",
    journal = "JHEP",
    volume = "02",
    pages = "087",
    year = "2020"
}

@article{Willenbrock:1987xz,
    author = "Willenbrock, Scott S. D.",
    title = "{Pair Production of $W$ and $Z$ Bosons and the Goldstone Boson Equivalence Theorem}",
    reportNumber = "MAD-PH-333",
    doi = "10.1016/S0003-4916(88)80016-2",
    journal = "Annals Phys.",
    volume = "186",
    pages = "15",
    year = "1988"
}

@article{Baglio:2018rcu,
    author = "Baglio, Julien and Ninh, Le Duc",
    title = "{Fiducial polarization observables in hadronic $WZ$ production: A next-to-leading order QCD+EW study}",
    eprint = "1810.11034",
    archivePrefix = "arXiv",
    primaryClass = "hep-ph",
    reportNumber = "IFIRSE-TH-2018-2",
    doi = "10.1007/JHEP04(2019)065",
    journal = "JHEP",
    volume = "04",
    pages = "065",
    year = "2019"
}

@article{ninh2023enhancing,
    author={Dao, Thi Nhung and Le, Duc Ninh},
    title = "{Enhancing the doubly-longitudinal polarization in $WZ$ production at the LHC}",
    eprint = "2302.03324",
    archivePrefix = "arXiv",
    primaryClass = "hep-ph",
    doi = "10.15625/0868-3166/18077",
    journal = "Commun. in Phys.",
    volume = "33",
    number = "3",
    pages = "223--234",
    year = "2023"
}

@article{BuarqueFranzosi:2019boy,
    author = "Franzosi, Diogo Buarque and Mattelaer, Olivier and Ruiz, Richard and Shil, Sujay",
    title = "{Automated predictions from polarized matrix elements}",
    eprint = "1912.01725",
    archivePrefix = "arXiv",
    primaryClass = "hep-ph",
    reportNumber = "CP3-19-52, VBSCan-PUB-09-19, MCnet-19-24, IP/BBSR/2019-10",
    doi = "10.1007/JHEP04(2020)082",
    journal = "JHEP",
    volume = "04",
    pages = "082",
    year = "2020"
}

@Article{Butterworth:2015oua,
      author         = "Butterworth, Jon and others",
      title          = "{PDF4LHC recommendations for LHC Run II}",
      journal        = "J. Phys. G",
      volume         = "43",
      year           = "2016",
      pages          = "023001",
      doi            = "10.1088/0954-3899/43/2/023001",
      eprint         = "1510.03865",
      archivePrefix  = "arXiv",
      primaryClass   = "hep-ph",
      reportNumber   = "OUTP-15-17P, SMU-HEP-15-12, TIF-UNIMI-2015-14,
                        LCTS-2015-27, CERN-PH-TH-2015-249",
      SLACcitation   = "%%CITATION = ARXIV:1510.03865;%%"
}

@article{Ball:2012cx,
      author         = "{NNPDF Collaboration} and Ball, Richard D. and others",
      title          = "{Parton distributions with LHC data}",
      journal        = "Nucl. Phys. B",
      volume         = "867",
      year           = "2013",
      pages          = "244",
      doi            = "10.1016/j.nuclphysb.2012.10.003",
      eprint         = "1207.1303",
      archivePrefix  = "arXiv",
      primaryClass   = "hep-ph",
      reportNumber   = "EDINBURGH-2012-08, IFUM-FT-997, FR-PHENO-2012-014,
                        RWTH-TTK-12-25, CERN-PH-TH-2012-037, SFB-CPP-12-47\,
                        --CERN-PH-TH-2012-037",
      SLACcitation   = "%%CITATION = ARXIV:1207.1303;%%"
}

@Article{Sjostrand:2007gs,
	author    = "Sj{\"o}strand, T. and Mrenna, S. and Skands, P.",
	title     = "{A brief introduction to PYTHIA 8.1}",
	journal   = "Comput. Phys. Commun.",
	volume    = "178",
	year      = "2008",
	pages     = "852-867",
	eprint    = "0710.3820",
	archivePrefix = "arXiv",
	primaryClass  =  "hep-ph",
	doi       = "10.1016/j.cpc.2008.01.036",
	SLACcitation  = "%%CITATION = 0710.3820;%%"
}

@article{Lonnblad:2001iq,
      author         = "L{\"o}nnblad, Leif",
      title          = "{Correcting the Colour-Dipole Cascade Model
                  with Fixed Order Matrix Elements}",
      journal        = "JHEP",
      volume         = "05",
      year           = "2002",
      pages          = "046",
      doi            = "10.1088/1126-6708/2002/05/046",
      eprint         = "hep-ph/0112284",
      archivePrefix  = "arXiv",
      XprimaryClass   = "hep-ph",
      reportNumber   = "LU-TP-01-38",
      SLACcitation   = "%%CITATION = HEP-PH/0112284;%%"
}

@Article{Campbell:1999ah,
      author         = "Campbell, John M. and Ellis, R. Keith",
      title          = "{Update on vector boson pair production at hadron colliders}",
      journal        = "Phys. Rev. D",
      volume         = "60",
      pages          = "113006",
      doi            = "10.1103/PhysRevD.60.113006",
      year           = "1999",
      eprint         = "hep-ph/9905386",
      archivePrefix  = "arXiv",
      reportNumber   = "FERMILAB-PUB-99-146-T",
      SLACcitation   = "%%CITATION = HEP-PH/9905386;%%",
}

@article{Agostinelli:2002hh,
      author         = "Agostinelli, S. and others",
      title          = "{\textsc{Geant4} -- a simulation toolkit}",
      journal        = "Nucl. Instrum. Meth. A",
      volume         = "506",
      year           = "2003",
      pages          = "250",
      doi            = "10.1016/S0168-9002(03)01368-8",
      reportNumber   = "SLAC-PUB-9350, FERMILAB-PUB-03-339",
      SLACcitation   = "%%CITATION = NUIMA,A506,250;%%"
}

@article{DAgostini:1995,
  author = "Giulio D'Agostini",
  title = "A multidimensional unfolding method based on Bayes' theorem",
  journal = "Nucl. Instrum. Meth. A",
  volume = "362",
  number = "2",
  pages = "487 - 498",
  year = "1995",
  issn = "0168-9002",
  doi = "10.1016/0168-9002(95)00274-X",
}

@Booklet{TMVA,
      author         = "Hoecker, Andreas and others",
      title          = "{TMVA - Toolkit for Multivariate Data Analysis}",
      year           = "2007",
      eprint         = "physics/0703039",
      archivePrefix  = "arXiv",
      primaryClass   = "physics.data-an",
}

@article{Zyla:2020zbs,
    author = "{Particle Data Group} and Zyla, P.A. and others",
    title = "{Review of Particle Physics}",
    journal = "PTEP",
    volume = "2020",
    number = "8",
    pages = "083C01",
    year = "2020",
    doi = "10.1093/ptep/ptaa104",
}

@Article{PERF-2007-01,
    author         = "{ATLAS Collaboration}",
    title          = "{The ATLAS Experiment at the CERN Large Hadron Collider}",
    journal        = "JINST",
    volume         = "3",
    year           = "2008",
    pages          = "S08003",
    doi            = "10.1088/1748-0221/3/08/S08003",
    primaryClass   = "hep-ex",
}

@Article{SOFT-2010-01,
    author         = "{ATLAS Collaboration}",
    title          = "{The ATLAS Simulation Infrastructure}",
    journal        = "Eur. Phys. J. C",
    volume         = "70",
    year           = "2010",
    pages          = "823",
    doi            = "10.1140/epjc/s10052-010-1429-9",
    eprint         = "1005.4568",
    archivePrefix  = "arXiv",
    primaryClass   = "physics.ins-det",
}

@Article{SUSY-2013-09,
    author         = "{ATLAS Collaboration}",
    title          = "{Search for supersymmetry at \(\sqrt{s} = 8\,\text{TeV}\) in final states with jets and two same-sign leptons or three leptons with the ATLAS detector}",
    journal        = "JHEP",
    volume         = "06",
    year           = "2014",
    pages          = "035",
    doi            = "10.1007/JHEP06(2014)035",
    reportNumber   = "CERN-PH-EP-2014-044",
    eprint         = "1404.2500",
    archivePrefix  = "arXiv",
    primaryClass   = "hep-ex",
}

@Article{PERF-2014-03,
    author         = "{ATLAS Collaboration}",
    title          = "{Performance of pile-up mitigation techniques for jets in \(pp\) collisions at \(\sqrt{s} = 8\,\text{TeV}\) using the ATLAS detector}",
    journal        = "Eur. Phys. J. C",
    volume         = "76",
    year           = "2016",
    pages          = "581",
    doi            = "10.1140/epjc/s10052-016-4395-z",
    reportNumber   = "CERN-PH-EP-2015-206",
    eprint         = "1510.03823",
    archivePrefix  = "arXiv",
    primaryClass   = "hep-ex",
}

@Article{PERF-2015-09,
    author         = "{ATLAS Collaboration}",
    title          = "{Jet reconstruction and performance using particle flow with the ATLAS Detector}",
    journal        = "Eur. Phys. J. C",
    volume         = "77",
    year           = "2017",
    pages          = "466",
    doi            = "10.1140/epjc/s10052-017-5031-2",
    reportNumber   = "CERN-EP-2017-024",
    eprint         = "1703.10485",
    archivePrefix  = "arXiv",
    primaryClass   = "hep-ex",
}

@Article{PERF-2016-06,
    author         = "{ATLAS Collaboration}",
    title          = "{Identification and rejection of pile-up jets at high pseudorapidity with the ATLAS detector}",
    journal        = "Eur. Phys. J. C",
    volume         = "77",
    year           = "2017",
    pages          = "580",
    doi            = "10.1140/epjc/s10052-017-5081-5",
    reportNumber   = "CERN-EP-2017-055",
    eprint         = "1705.02211",
    archivePrefix  = "arXiv",
    primaryClass   = "hep-ex",
    related        = "PERF-2016-06-err",
    relatedstring  = "Erratum:",
}

@Article{PERF-2016-07,
    author         = "{ATLAS Collaboration}",
    title          = "{Performance of missing transverse momentum reconstruction with the ATLAS detector using proton--proton collisions at \(\sqrt{s} = 13\,\text{TeV}\)}",
    journal        = "Eur. Phys. J. C",
    volume         = "78",
    year           = "2018",
    pages          = "903",
    doi            = "10.1140/epjc/s10052-018-6288-9",
    reportNumber   = "CERN-EP-2017-274",
    eprint         = "1802.08168",
    archivePrefix  = "arXiv",
    primaryClass   = "hep-ex",
}

@Article{TOPQ-2016-11,
    author         = "{ATLAS Collaboration}",
    title          = "{Measurement of the \(t\bar{t}Z\) and \(t\bar{t}W\) cross sections in proton--proton collisions at \(\sqrt{s} = 13\,\text{TeV}\) with the ATLAS detector}",
    journal        = "Phys. Rev. D",
    volume         = "99",
    year           = "2019",
    pages          = "072009",
    doi            = "10.1103/PhysRevD.99.072009",
    reportNumber   = "CERN-EP-2018-331",
    eprint         = "1901.03584",
    archivePrefix  = "arXiv",
    primaryClass   = "hep-ex",
}

@Article{TRIG-2016-01,
    author         = "{ATLAS Collaboration}",
    title          = "{Performance of the ATLAS trigger system in 2015}",
    journal        = "Eur. Phys. J. C",
    volume         = "77",
    year           = "2017",
    pages          = "317",
    doi            = "10.1140/epjc/s10052-017-4852-3",
    reportNumber   = "CERN-EP-2016-241",
    eprint         = "1611.09661",
    archivePrefix  = "arXiv",
    primaryClass   = "hep-ex",
}

@Article{HIGG-2017-02,
    author         = "{ATLAS Collaboration}",
    title          = "{Evidence for the associated production of the Higgs boson and a top quark pair with the ATLAS detector}",
    journal        = "Phys. Rev. D",
    volume         = "97",
    year           = "2018",
    pages          = "072003",
    doi            = "10.1103/PhysRevD.97.072003",
    reportNumber   = "CERN-EP-2017-281",
    eprint         = "1712.08891",
    archivePrefix  = "arXiv",
    primaryClass   = "hep-ex",
}

@Article{PERF-2017-01,
    author         = "{ATLAS Collaboration}",
    title          = "{Electron reconstruction and identification in the ATLAS experiment using the 2015 and 2016 LHC proton--proton collision data at \(\sqrt{s} = 13\,\text{TeV}\)}",
    journal        = "Eur. Phys. J. C",
    volume         = "79",
    year           = "2019",
    pages          = "639",
    doi            = "10.1140/epjc/s10052-019-7140-6",
    reportNumber   = "CERN-EP-2018-273",
    eprint         = "1902.04655",
    archivePrefix  = "arXiv",
    primaryClass   = "physics.ins-det",
}

@Article{PERF-2017-03,
    author         = "{ATLAS Collaboration}",
    title          = "{Electron and photon energy calibration with the ATLAS detector using 2015--2016 LHC proton--proton collision data}",
    journal        = "JINST",
    volume         = "14",
    year           = "2019",
    pages          = "P03017",
    doi            = "10.1088/1748-0221/14/03/P03017",
    reportNumber   = "CERN-EP-2018-296",
    eprint         = "1812.03848",
    archivePrefix  = "arXiv",
    primaryClass   = "hep-ex",
}

@Article{STDM-2017-06,
    author         = "{ATLAS Collaboration}",
    title          = "{Observation of Electroweak Production of a Same-Sign \(W\) Boson Pair in Association with Two Jets in \(pp\) Collisions at \(\sqrt{s} = 13\,\text{TeV}\) with the ATLAS Detector}",
    journal        = "Phys. Rev. Lett.",
    volume         = "123",
    year           = "2019",
    pages          = "161801",
    doi            = "10.1103/PhysRevLett.123.161801",
    reportNumber   = "CERN-EP-2019-008",
    eprint         = "1906.03203",
    archivePrefix  = "arXiv",
    primaryClass   = "hep-ex",
}

@Article{STDM-2017-22,
    author         = "{ATLAS Collaboration}",
    title          = "{Evidence for the production of three massive vector bosons with the ATLAS detector}",
    journal        = "Phys. Lett. B",
    volume         = "798",
    year           = "2019",
    pages          = "134913",
    doi            = "10.1016/j.physletb.2019.134913",
    reportNumber   = "CERN-EP-2019-041",
    eprint         = "1903.10415",
    archivePrefix  = "arXiv",
    primaryClass   = "hep-ex",
}

@Article{EGAM-2018-01,
    author         = "{ATLAS Collaboration}",
    title          = "{Electron and photon performance measurements with the ATLAS detector using the 2015--2017 LHC proton--proton collision data}",
    journal        = "JINST",
    volume         = "14",
    year           = "2019",
    pages          = "P12006",
    doi            = "10.1088/1748-0221/14/12/P12006",
    reportNumber   = "CERN-EP-2019-145",
    eprint         = "1908.00005",
    archivePrefix  = "arXiv",
    primaryClass   = "hep-ex",
}

@Article{FTAG-2018-01,
    author         = "{ATLAS Collaboration}",
    title          = "{ATLAS \(b\)-jet identification performance and efficiency measurement with \(t\bar{t}\) events in \(pp\) collisions at \(\sqrt{s} = 13\,\text{TeV}\)}",
    journal        = "Eur. Phys. J. C",
    volume         = "79",
    year           = "2019",
    pages          = "970",
    doi            = "10.1140/epjc/s10052-019-7450-8",
    reportNumber   = "CERN-EP-2019-132",
    eprint         = "1907.05120",
    archivePrefix  = "arXiv",
    primaryClass   = "hep-ex",
}

@Article{JETM-2018-05,
    author         = "{ATLAS Collaboration}",
    title          = "{Jet energy scale and resolution measured in proton--proton collisions at \(\sqrt{s} = 13\,\text{TeV}\) with the ATLAS detector}",
    journal        = "Eur. Phys. J. C",
    volume         = "81",
    year           = "2021",
    pages          = "689",
    doi            = "10.1140/epjc/s10052-021-09402-3",
    reportNumber   = "CERN-EP-2020-083",
    eprint         = "2007.02645",
    archivePrefix  = "arXiv",
    primaryClass   = "hep-ex",
}

@Article{MUON-2018-03,
    author         = "{ATLAS Collaboration}",
    title          = "{Muon reconstruction and identification efficiency in ATLAS using the full Run~2 \(pp\) collision data set at \(\sqrt{s} = 13\,\text{TeV}\)}",
    journal        = "Eur. Phys. J. C",
    volume         = "81",
    year           = "2021",
    pages          = "578",
    doi            = "10.1140/epjc/s10052-021-09233-2",
    reportNumber   = "CERN-EP-2020-199",
    eprint         = "2012.00578",
    archivePrefix  = "arXiv",
    primaryClass   = "hep-ex",
}

@Article{STDM-2018-03,
    author         = "{ATLAS Collaboration}",
    title          = "{Measurement of \(W^{\pm}Z\) production cross sections and gauge boson polarisation in \(pp\) collisions at \(\sqrt{s} = 13\,\text{TeV}\) with the ATLAS detector}",
    journal        = "Eur. Phys. J. C",
    volume         = "79",
    year           = "2019",
    pages          = "535",
    doi            = "10.1140/epjc/s10052-019-7027-6",
    reportNumber   = "CERN-EP-2018-327",
    eprint         = "1902.05759",
    archivePrefix  = "arXiv",
    primaryClass   = "hep-ex",
}

@Article{TOPQ-2018-01,
    author         = "{ATLAS Collaboration}",
    title          = "{Observation of the associated production of a top quark and a \(Z\) boson in \(pp\) collisions at \(\sqrt{s} = 13\,\text{TeV}\) with the ATLAS detector}",
    journal        = "JHEP",
    volume         = "07",
    year           = "2020",
    pages          = "124",
    doi            = "10.1007/JHEP07(2020)124",
    reportNumber   = "CERN-EP-2019-273",
    eprint         = "2002.07546",
    archivePrefix  = "arXiv",
    primaryClass   = "hep-ex",
}

@Article{TRIG-2018-05,
    author         = "{ATLAS Collaboration}",
    title          = "{Performance of electron and photon triggers in ATLAS during LHC Run~2}",
    journal        = "Eur. Phys. J. C",
    volume         = "80",
    year           = "2020",
    pages          = "47",
    doi            = "10.1140/epjc/s10052-019-7500-2",
    reportNumber   = "CERN-EP-2019-169",
    eprint         = "1909.00761",
    archivePrefix  = "arXiv",
    primaryClass   = "hep-ex",
}

@Article{STDM-2021-05,
    author         = "{ATLAS Collaboration}",
    title          = "{Evidence of pair production of longitudinally polarised vector bosons and study of CP properties in \(ZZ \to 4\ell\) events with the ATLAS detector at \(\sqrt{s} = 13\,\text{TeV}\)}",
    journal        = "JHEP",
    volume         = "12",
    year           = "2023",
    pages          = "107",
    doi            = "10.1007/JHEP12(2023)107",
    reportNumber   = "CERN-EP-2023-199",
    eprint         = "2310.04350",
    archivePrefix  = "arXiv",
    primaryClass   = "hep-ex",
}

@Article{STDM-2022-01,
    author         = "{ATLAS Collaboration}",
    title          = "{Observation of gauge boson joint-polarisation states in \(W^{\pm}Z\) production from \(pp\) collisions at \(\sqrt{s} = 13\,\text{TeV}\) with the ATLAS detector}",
    journal        = "Phys. Lett. B",
    volume         = "843",
    year           = "2023",
    pages          = "137895",
    doi            = "10.1016/j.physletb.2023.137895",
    reportNumber   = "CERN-EP-2022-202",
    eprint         = "2211.09435",
    archivePrefix  = "arXiv",
    primaryClass   = "hep-ex",
}

@Booklet{ATL-SOFT-PUB-2021-001,
    author         = "{ATLAS Collaboration}",
    title          = "{The ATLAS Collaboration Software and Firmware}",
    howpublished   = "{ATL-SOFT-PUB-2021-001}",
    url            = "https://cds.cern.ch/record/2767187",
    year           = "2021",
}

@Booklet{ATL-SOFT-PUB-2023-001,
    author         = "{ATLAS Collaboration}",
    title          = "{ATLAS Computing Acknowledgements}",
    howpublished   = "{ATL-SOFT-PUB-2023-001}",
    url            = "https://cds.cern.ch/record/2869272",
    year           = "2023",
}

@Article{CMS-EWK-11-009,
    author         = "{CMS Collaboration}",
    title          = "{Measurement of the \(W \gamma\) and \(Z \gamma\) inclusive cross sections in \(pp\) collisions at \(\sqrt{s} = 7\,\text{TeV}\) and limits on anomalous triple gauge boson couplings}",
    journal        = "Phys. Rev. D",
    volume         = "89",
    year           = "2014",
    pages          = "092005",
    doi            = "10.1103/PhysRevD.89.092005",
    reportNumber   = "CERN-PH-EP-2013-108",
    eprint         = "1308.6832",
    archivePrefix  = "arXiv",
    primaryClass   = "hep-ex",
}

@Article{CMS-SMP-20-005,
    author         = "{CMS Collaboration}",
    title          = "{Measurement of \(W^{\pm}\gamma\) differential cross sections in proton--proton collisions at \(\sqrt{s} = 13\,\text{TeV}\) and effective field theory constraints}",
    journal        = "Phys. Rev. D",
    volume         = "105",
    year           = "2022",
    pages          = "052003",
    doi            = "10.1103/PhysRevD.105.052003",
    reportNumber   = "CERN-EP-2021-219",
    eprint         = "2111.13948",
    archivePrefix  = "arXiv",
    primaryClass   = "hep-ex",
}

@Booklet{ATLAS-CONF-2014-018,
    author         = "{ATLAS Collaboration}",
    title          = "{Tagging and suppression of pileup jets with the ATLAS detector}",
    howpublished   = "{ATLAS-CONF-2014-018}",
    url            = "https://cds.cern.ch/record/1700870",
    year           = "2014",
}

@Booklet{ATLAS-CONF-2019-021,
    author         = "{ATLAS Collaboration}",
    title          = "{Luminosity determination in \(pp\) collisions at \(\sqrt{s} = 13\,\text{TeV}\) using the ATLAS detector at the LHC}",
    howpublished   = "{ATLAS-CONF-2019-021}",
    url            = "https://cds.cern.ch/record/2677054",
    year           = "2019",
}

@Booklet{ATL-PHYS-PUB-2016-002,
    author         = "{ATLAS Collaboration}",
    title          = "{Multi-boson simulation for \(13~\text{TeV}\) ATLAS analyses}",
    howpublished   = "{ATL-PHYS-PUB-2016-002}",
    url            = "https://cds.cern.ch/record/2119986",
    year           = "2016",
}

@Booklet{ATL-DAQ-PUB-2019-001,
    author         = "{ATLAS Collaboration}",
    title          = "{Trigger Menu in 2018}",
    howpublished   = "{ATL-DAQ-PUB-2019-001}",
    url            = "https://cds.cern.ch/record/2693402",
    year           = "2019",
}
 
\clearpage
 
\begin{flushleft}
\hypersetup{urlcolor=black}
{\Large The ATLAS Collaboration}

\bigskip

\AtlasOrcid[0000-0002-6665-4934]{G.~Aad}$^\textrm{\scriptsize 103}$,
\AtlasOrcid[0000-0001-7616-1554]{E.~Aakvaag}$^\textrm{\scriptsize 16}$,
\AtlasOrcid[0000-0002-5888-2734]{B.~Abbott}$^\textrm{\scriptsize 121}$,
\AtlasOrcid[0000-0002-1002-1652]{K.~Abeling}$^\textrm{\scriptsize 55}$,
\AtlasOrcid[0000-0001-5763-2760]{N.J.~Abicht}$^\textrm{\scriptsize 49}$,
\AtlasOrcid[0000-0002-8496-9294]{S.H.~Abidi}$^\textrm{\scriptsize 29}$,
\AtlasOrcid[0009-0003-6578-220X]{M.~Aboelela}$^\textrm{\scriptsize 44}$,
\AtlasOrcid[0000-0002-9987-2292]{A.~Aboulhorma}$^\textrm{\scriptsize 35e}$,
\AtlasOrcid[0000-0001-5329-6640]{H.~Abramowicz}$^\textrm{\scriptsize 152}$,
\AtlasOrcid[0000-0002-1599-2896]{H.~Abreu}$^\textrm{\scriptsize 151}$
% ...
\bigskip
\\

% ...
$^{16}$Department for Physics and Technology, University of Bergen, Bergen; Norway.\\
% ...
$^{29}$Physics Department, Brookhaven National Laboratory, Upton NY; United States of America.\\
% ...
$^{35}$$^{(a)}$Facult\'e des Sciences Ain Chock, R\'eseau Universitaire de Physique des Hautes Energies - Universit\'e Hassan II, Casablanca;$^{(b)}$Facult\'{e} des Sciences, Universit\'{e} Ibn-Tofail, K\'{e}nitra;$^{(c)}$Facult\'e des Sciences Semlalia, Universit\'e Cadi Ayyad, LPHEA-Marrakech;$^{(d)}$LPMR, Facult\'e des Sciences, Universit\'e Mohamed Premier, Oujda;$^{(e)}$Facult\'e des sciences, Universit\'e Mohammed V, Rabat;$^{(f)}$Institute of Applied Physics, Mohammed VI Polytechnic University, Ben Guerir; Morocco.\\
% ...
$^{44}$Physics Department, Southern Methodist University, Dallas TX; United States of America.\\
% ...
$^{49}$Fakult\"{a}t Physik , Technische Universit{\"a}t Dortmund, Dortmund; Germany.\\
% ...
$^{55}$II. Physikalisches Institut, Georg-August-Universit\"{a}t G\"ottingen, G\"ottingen; Germany.\\
% ...
$^{103}$CPPM, Aix-Marseille Universit\'e, CNRS/IN2P3, Marseille; France.\\
% ...
$^{121}$Homer L. Dodge Department of Physics and Astronomy, University of Oklahoma, Norman OK; United States of America.\\
% ...
$^{151}$Department of Physics, Technion, Israel Institute of Technology, Haifa; Israel.\\
$^{152}$Raymond and Beverly Sackler School of Physics and Astronomy, Tel Aviv University, Tel Aviv; Israel.\\
% ...
$^{a}$ Also Affiliated with an institute covered by a cooperation agreement with CERN.\\
$^{b}$ Also at An-Najah National University, Nablus; Palestine.\\
$^{c}$ Also at Borough of Manhattan Community College, City University of New York, New York NY; United States of America.\\
$^{d}$ Also at Center for High Energy Physics, Peking University; China.\\
$^{e}$ Also at Center for Interdisciplinary Research and Innovation (CIRI-AUTH), Thessaloniki; Greece.\\
$^{f}$ Also at Centro Studi e Ricerche Enrico Fermi; Italy.\\
% ...

\end{flushleft}


\end{document}